\documentclass[aps,pra, twocolumn,superscriptaddress,longbibliography]{revtex4-2}
\usepackage{graphicx}
\usepackage{float}

\usepackage{derivative}
\usepackage{xcolor}
\usepackage[margin=0.68in]{geometry}
\usepackage{siunitx}
\usepackage{soul}
\usepackage{minitoc}
\usepackage{color}
\usepackage{amssymb}
\usepackage{amsmath, scalerel, breqn}
\usepackage{physics}

\definecolor{dark-red}{rgb}{0.4,0.15,0.15}
\definecolor{dark-blue}{rgb}{0.15,0.15,0.4}
\definecolor{medium-blue}{rgb}{0,0,0.5}
\usepackage{hyperref}
\hypersetup{
    colorlinks, linkcolor={dark-red},
    citecolor={dark-blue}, urlcolor={medium-blue}
}

\newcommand{\FSIM}{\text{fSim}}

\begin{document}
\doparttoc 
\faketableofcontents 
\part{} 

\setcitestyle{super}

\title{Formation of robust bound states of interacting photons}
\author{Google Quantum AI and Collaborators}
\date{\today}

\begin{abstract}
Systems of correlated particles appear in many fields of science and represent some of the most intractable puzzles in nature. The computational challenge in these systems arises when interactions become comparable to other energy scales, which makes the state of each particle depend on all other particles\,\cite{Eckle2019}. The lack of general solutions for the 3-body problem and acceptable theory for strongly correlated electrons shows that our understanding of correlated systems fades when the particle number or the interaction strength increases. One of the hallmarks of interacting systems is the formation of multi-particle bound states\,\cite{Bethe1931,E8_2010, ganahl_2012,Bloch2013,Efimov1970,Kraemer2006,Kunitski2015,GreeneRMP2017}. In a ring of 24 superconducting qubits, we develop a high fidelity parameterizable fSim gate that we use to implement the periodic quantum circuit of the spin-1/2 XXZ model, an archetypal model of interaction. By placing  microwave photons in adjacent qubit sites, we study the propagation of these excitations and observe their bound nature for up to 5 photons. We devise a phase sensitive method for constructing the few-body spectrum of the bound states and extract their pseudo-charge by introducing a synthetic flux. By introducing interactions between the ring and additional qubits, we observe an unexpected resilience of the bound states to integrability breaking. This finding goes against the common wisdom that bound states in non-integrable systems are unstable when their energies overlap with the continuum spectrum. Our work provides experimental evidence for bound states of interacting photons and discovers their stability beyond the integrability limit.
\end{abstract}

\maketitle
Photons that propagate in vacuum do not interact with each other; however, many technological applications and the study of fundamental physics require interacting photons. 
Consequently, realizing quantum platforms 
with strong interactions between photons constitutes a major scientific goal\,\cite{Kimble2005,Vuletic2014}. In this regard, superconducting circuits are promising candidates since they provide a configurable lattice where microwave photons can be confined to a qubit site, hop between the sites, and interact with each other. Each site can host localized electromagnetic oscillations and hence be occupied with a discrete number of microwave photon excitations. The tunability of coupling elements allows photons to hop between the sites, and the non-linearity of qubits leads to interaction between the photons. The zero- and single-photon occupancies of qubits are used as the $\ket{0}$ and $\ket{1}$ states in quantum information processing. Here we also confine the dynamics to zero or single occupancy for a given qubit, the so-called hard core boson limit, and show that microwave photons can 
remain adjacent and form coherent bound states.

The advent of quantum processors is giving rise to a paradigm shift in the studies of correlated systems\,\cite{Blatt_NatPhys_2012,Gross_Science_2017,Carusotto_NatPhys_2020,yao2022,Knap2022}. While theoretical studies of condensed matter models were focused on Hamiltonian systems for many decades, high-fidelity quantum processors commonly operate based on unitary gates rather than continuous Hamiltonian dynamics. This experimental access to periodic (Floquet) unitary dynamics opens the door to a plethora of non-equilibrium phenomena\,\cite{Vedika2016}. Since such periodic dynamics often cannot be described in terms of a local Hamiltonian, established results are fewer and far in between\,\cite{Shirley1965, ZELDOVICH1966, Sambe1973}. For instance, until recently, there was no theoretically known example of bound state formation for interacting Floquet dynamics. 

Integrable models form the cornerstone of our understanding of dynamical systems and can serve to benchmark quantum processors. A relevant example of an interacting integrable model is the 1D quantum spin-$1/2$ XXZ model which is known to support bound states\,\cite{Bethe1931,Fonseca2006, E8_2010, ganahl_2012,Bloch2013}. Recently, the shared symmetries of the spin-$1/2$ XXZ Hamiltonian model with its Floquet counterpart led to a proof for the integrability of the XXZ Floquet quantum circuits\,\cite{Claeys2022, Prosen2019}. Later, Aleiner obtained the full spectrum for these Floquet systems and provided analytical results for bound states\,\cite{Aleiner2021}. The advantage of using quantum processors in studying these models becomes apparent when going beyond the integrability limit, where the classical counterpart of the circuit shows chaos, and analytical and numerical techniques fail to scale favorably. 

\begin{figure}[h!]
    \centering
    \includegraphics[width=0.48\textwidth]{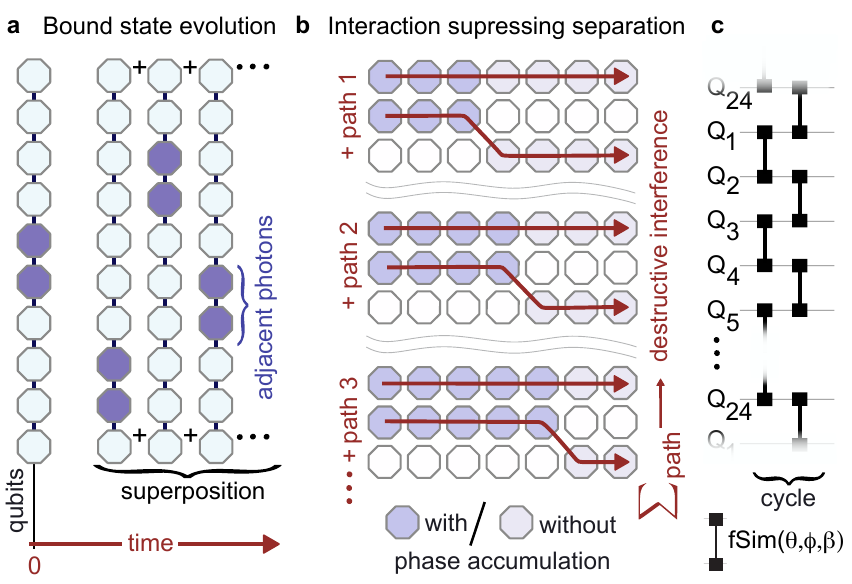} 
    \caption{\small{\textbf{Bound states of photons. a,} In a 1D chain of qubits hosting bound states, an initial state with adjacent photons evolves into a superposition of states in which the photons remain bound together. \textbf{b,} Interactions between photons can lead to destructive interference for paths in which photons do not stay together, thus suppressing separation. \textbf{c,} Schematic of the gate sequence used in this work. Each cycle of evolution contains two layers of fSim gates that connect the even and odd pairs respectively. The fSim gate has three controllable parameters that set the kinetic energy ($\theta$), the interaction strength ($\phi$) and a synthetic magnetic flux ($\beta$). The median gate fidelity, measured with cross-entropy benchmarking, is $1.1\%$ (see supplementary information).}}
    \label{fig:fig1}
\end{figure}

To define systems with bound states, consider a chain of coupled qubits and the unitary evolution $\hat{U}$ of interacting photons on this array. We divide the computational space of all bitstrings with $n_{\mathrm{ph}}$ photons into two sets: one set $\mathcal{T}$ composed of all bitstrings in which all photons are in adjacent sites, e.g. $\ket{00...011100...00}$; the other set $\mathcal{S}$ contains all other $n_{\mathrm{ph}}$ bitstrings, e.g. $\ket{00...101001...00}$. A bound state is formed when the eigenstates of the system can be expanded as the superposition of bitstrings mainly in $\mathcal{T}$ and with smaller weight in $\mathcal{S}$. Therefore, for any initial state $\ket{\psi_0} \in \mathcal{T}$ the photons remain adjacent at all future times $\ket{\psi}= \hat{U} \ket{\psi_0}$, which implies that almost every projective measurement returns a bitstring in $\mathcal{T}$\,(Fig.\,\ref{fig:fig1}a). 

The emergence of a thermodynamic phase or the formation of a bound state in Floquet dynamics seems rather implausible at first sight. In a closed Floquet system there is no notion of lowest energy, a key concept in equilibrium physics. Therefore, the energy minimization that commonly 
stabilizes bound states in e.g. atoms does not hold. In the absence of interactions and in 1D, photons hop independently and the evolution can be mapped to that of free fermions. In this limit, obviously, no bound state can be formed. The key question of bound state formation is whether the effect of kinetic energy\,(hopping) that moves photons away from each other could be balanced by interactions. In Fig.\,\ref{fig:fig1}b, we provide a plausibility argument to illustrate this point. Consider two photons that are initially occupying adjacent sites, in the low kinetic energy regime where maximum one hopping event occurs in the span of a few cycles. In the spirit of Feynman path formulation, the probability of a given configuration at a later time can be obtained from summing over all possible paths that lead to that configuration with proper weights. When photons are in adjacent sites, they accumulate phase due to the interaction. In the three depicted paths, the accumulated phases are different, thus leading to destructive interference. Hence, the interactions suppress the probability of unbound configurations and facilitate the formation of bound states. 

The control sequence used to generate unitary evolution in our experiment consists of a periodic application of entangling gates in a 1D ring of $N_Q=24$ qubits\,(Fig.\,1c). Within each cycle, 2-qubit fSim gates are applied between all pairs in the ring. In the 2-qubit subspace, $\{\ket{00},\ket{01},\ket{10},\ket{11}\}$, this gate can be written as
\begin{equation}
    \FSIM(\theta, \phi, \beta) = \begin{pmatrix} 1 & 0 & 0 & 0 \\
                                         0 & \cos \theta & ie^{i \beta} \sin \theta & 0 \\
                                         0 & i e^{-i\beta}\sin \theta & \cos \theta & 0 \\
                                         0 & 0 & 0 & e^{i\phi} \\
    \end{pmatrix},
\end{equation}
\noindent
where $\theta$ and $\beta$ set the amplitude and phase, respectively, of hopping between adjacent qubit lattice sites, and the conditional-phase angle $\phi$ imparts a phase on the $\ket{11}$ state upon interaction of two adjacent photons. In the supplementary information, we show that we can achieve this gate with high fidelity ($\sim1\%$) for several angles. In the following, we will denote $\FSIM(\theta, \phi,\beta=0)$ as $\FSIM(\theta, \phi)$. The qubit chain is periodically driven by a quantum circuit, with the cycle unitary:
\begin{equation}
    \hat{U}_F= {\prod_{\text{even bonds}}\FSIM(\theta, \phi,\beta) \prod_{\text{odd bonds}}\FSIM(\theta, \phi,\beta)}. 
\end{equation}
\noindent
In the limit of $\beta=0$ and $\theta, \phi \rightarrow 0$, this model becomes the Trotter-Suzuki expansion of the XXZ Hamiltonian model.

\begin{figure*}
    \centering
    \includegraphics[width=\textwidth]{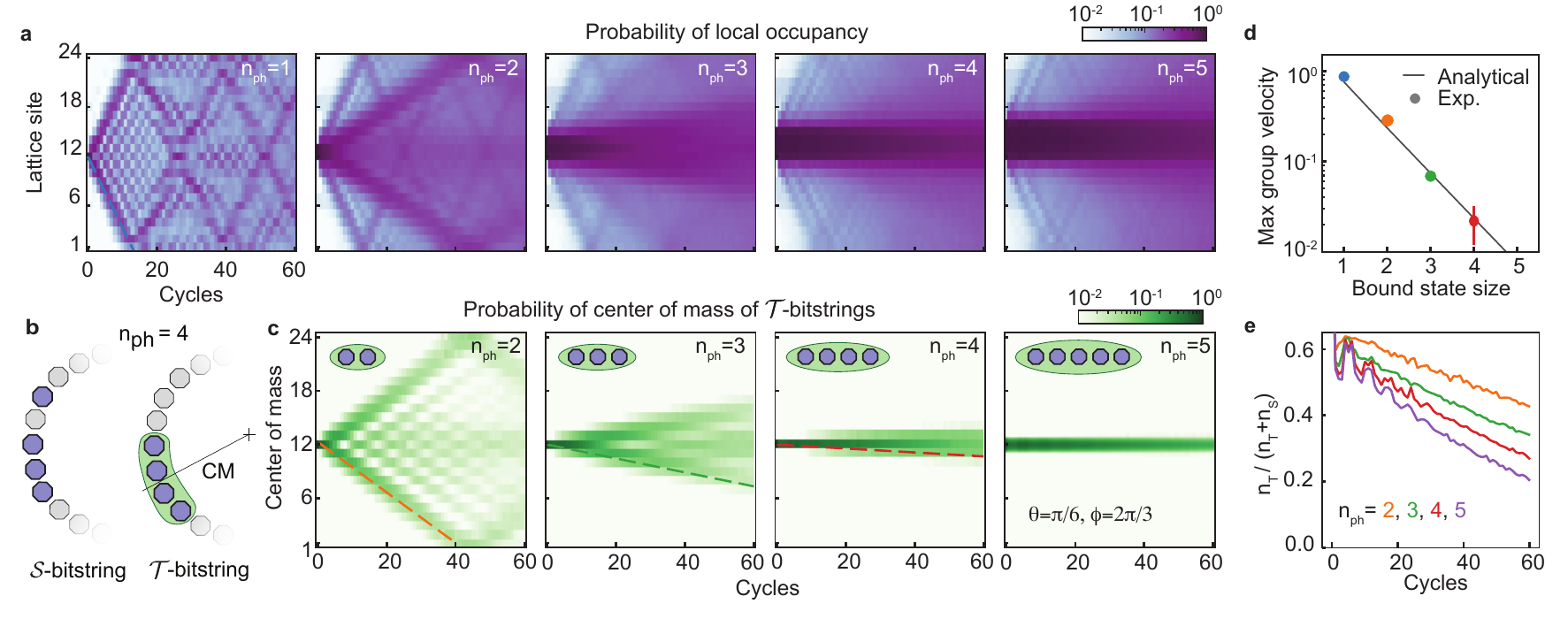}
    \caption{\textbf{Trajectory of bound photons. a}, Time- and site-resolved photon occupancy on a 24-qubits ring for photon numbers $n_{\mathrm{ph}}=1-5$. To measure a $n_{\mathrm{ph}}$-photon bound state, $n_{\mathrm{ph}}$ adjacent qubits are prepared in the $\ket{1}$ state. \textbf{b}, Schematic and example of bitstrings in $\mathcal{T}$ and $\mathcal{S}$. Center of mass is defined as the center of $n_{\mathrm{ph}}$ adjacent occupied sites. \textbf{c}, Evolution of the center of mass of $n_{\mathrm{ph}}$-bound states. Each trajectory is similar to the single photon case, highlighting the composite nature of the bound states. \textbf{d}, Extracted maximum group velocity from the trajectory of the center of mass. Black line: theoretical prediction. \textbf{e}, Decay of the bound state into the single excitations continuum due to dephasing. For all panels, $\theta = \pi/6$ and $\phi=2\pi/3$, and the trajectories are averaged over all possible initial states. Data are post-selected for number of excitations equal to $n_{\mathrm{ph}}$.}
    \label{fig:fig2}
\end{figure*}

To quantify to what extent photons remain bound together, we prepare an initial state with $n_{\mathrm{ph}}$ photons at adjacent sites and measure the photon occupancy of all sites after each cycle with approximately 5,000 repetitions. In Fig.\,2a we plot the average photon occupancy $(1-\langle \hat{Z_j} \rangle)/2$ on each site $j$ as a function of circuit depth for the fSim angles $\theta=\pi/6$ and $\phi=2\pi/3$. Since the fSim gates are excitation number conserving, all data are post-selected for the bitstrings with the proper number of excitations, which allows us to mitigate errors induced by population decay. While $n_{\mathrm{ph}}=1$ is not a bound state, it provides a benchmark, where we can clearly see the quantum random walk of a single particle and its familiar interference pattern. For $n_{\mathrm{ph}}=2$, we observe the appearance of two wavefronts: the fastest one corresponds to unbound photons, whereas the other one corresponds to the 2-photon bound state. For $n_{\mathrm{ph}}>2$, the concentration of the population near the center indicates that the photons do not disperse far, but instead stay close to each other. In the supplementary information, we also present the situation where the initial photons are not adjacent, in which case the system tends toward a uniform distribution.  

To extract the wavefront velocity, we select the measured bitstrings in which the photons remain adjacent, i.e. in $\mathcal{T}$, and discard the ones in $\mathcal{S}$. In panel c, we present the spatially and temporally resolved probabilities of the ``center of photon mass'' (CM, Fig.\,2b) of these $\mathcal{T}$ bitstrings. With this selection, the first panel in c shows a very similar pattern to the single-particle propagation in a, highlighting the composite nature of the bound state. The propagation velocities of the bound states can now be easily seen, and as expected, the larger bound states propagate more slowly. The wavefronts propagate with constant velocity, indicating that the bound photons move ballistically and without effects of impurity scattering. The extracted maximum group velocities of the bound states, $v^{\text{max}}_{\text{g}}$ (Fig.\,2d), match very well with that corresponding to the analytical dispersion relations derived in ref \cite{Aleiner2021}, which take the same functional form for all $n_{\mathrm{ph}}$:
\begin{equation}\label{eq:dispersion_relation}
    \cos(E(k) - \chi) = \cos^2(\alpha) - \sin^2(\alpha) \cos(k),   
\end{equation}
\noindent
where $\alpha$ and $\chi$ are functions of $n_{\mathrm{ph}}$, $\theta$, and $\phi$\ (see supplementary information for exact forms).

In order to characterize the stability of the bound state, it is useful to consider the evolution of the fraction of bitstrings in which the photons remain adjacent, $n_{\mathcal{T}}/(n_{\mathcal{T}} + n_{\mathcal{S}})$ (where $n_{\mathcal{T}(\mathcal{S})}$ is the number of bitstrings in $\mathcal{T}(\mathcal{S})$),  
which reflects contributions from both internal unitary dynamics as well as external decoherence\,(Fig.\,2e). In the absence of dephasing, $n_{\mathcal{T}}$ should reach a steady-state value after the observed initial drop. However, we observe a slow decay which we attribute to the dephasing of the qubits, since the data is post-selected to remove $T_1$ photon loss effects. A remarkable feature of the data is that the decay rate for various $n_{\mathrm{ph}}$ values is the same, indicating that this decay is dominated by bond breaking at the edges of the bound state. 

\begin{figure*}
    \centering
  \includegraphics[width=\textwidth]{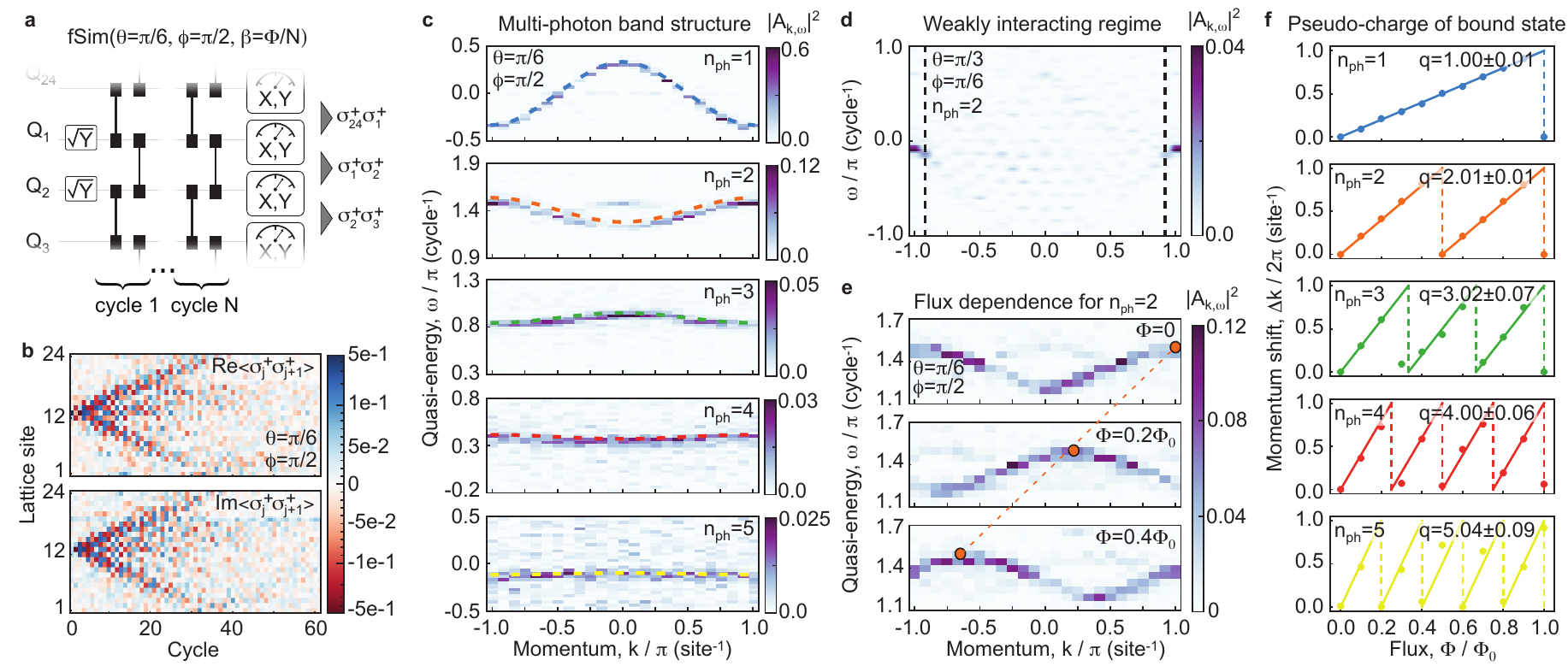}
    \caption{\textbf{Band structure of multi-photon bound states. a}, Schematic of circuit used for many-body spectroscopy. $n_{\mathrm{ph}}$ adjacent qubits are prepared in the $\ket{+}$-state, before evolving the state with a variable number of fSim gates. The phase of the bound state is probed by measuring the correlator $\langle \sigma_i^+..\sigma_{i+n_{\mathrm{ph}}-1}^+\rangle$ for all sets of $n_{\mathrm{ph}}$ adjacent qubits. \textbf{b}, Real (top) and imaginary (bottom) parts of the $n_{\mathrm{ph}}=2$ correlator. \textbf{c}, Band structure for $n_{\mathrm{ph}}=1-5$ (top to bottom), obtained via a 2D Fourier transform in space and time of the $n_{\mathrm{ph}}$-correlators. Color scale: absolute square of the Fourier transform, $|A_{k,\omega}|^2$. Dashed curves: theoretical prediction in Eqn. \ref{eq:dispersion_relation}. \textbf{d}, Band structure for $n_{\mathrm{ph}}=2$ in the weakly interacting ($\phi<2\theta$) regime, displaying the emergence of a bound state only at momenta near $k=\pm\pi$. Dashed black lines: theoretically predicted momentum threshold for the existence of the bound state (see supplementary information). \textbf{e}, Flux dependence of the $n_{\mathrm{ph}}=2$ band structure, displaying a gradual momentum shift as the flux increases ($\Phi_0=2\pi N_Q$). Orange circles and dashed line indicate the peak position of the band. \textbf{f}, Extracted momentum shifts as a function of flux for $n_{\mathrm{ph}}=1-5$ (top to bottom), indicating that the rate of shifting scales linearly with the photon number of the bound states, \textit{i.e.} the pseudo-charge $q$ of each bound state is proportional to its number of photons. Colored lines: theoretical prediction.} 
    \label{fig:fig3}
\end{figure*}

To show that the bound photons are quasiparticles with well-defined momentum, energy, and charge, we study the spectrum of the bound states using a many-body spectroscopy technique \cite{roushan2017spectroscopic}. We measure the energy of the bound states by comparing their accumulated phase over time relative to the vacuum state $\ket{0}^{\otimes N_Q}$. This is achieved by preparing $n_{\mathrm{ph}}$ adjacent qubits in the $\ket{+X}$-state and measuring the following $n_{\mathrm{ph}}$-body correlator that couples the bound states with the vacuum state:
\begin{equation}
    \langle C_{j,n_{\mathrm{ph}}}\rangle=\langle\Pi_{i=j}^{j+n_{\mathrm{ph}}-1}\sigma_{i}^+\rangle=\langle\Pi_{i=j}^{j+n_{\mathrm{ph}}-1}(X_i+iY_i)\rangle
\end{equation}
\noindent
for all sets of $n_{\mathrm{ph}}$ adjacent qubits (Fig.\,3a). This protocol is based on measuring the Green function of the system. While the correlator above is not Hermitian, it can be reconstructed by measuring its constituent terms (e.g. $\langle X_jX_{j+1}\rangle-\langle Y_jY_{j+1}\rangle+i\langle X_jY_{j+1}\rangle+i\langle Y_jX_{j+1}\rangle$ for $n_{\mathrm{ph}}=2$) and summing these with the proper complex pre-factors. We note that since  $C_{j,n_{\mathrm{ph}}}$ only couples the $n_{\mathrm{ph}}$-photon terms to the vacuum, the initial product state used here serves the same purpose as an entangled superposition state $\ket{000..00}+\ket{00..0110..00}$. By expanding these states in the momentum basis\,($k$-space), it becomes evident that $\langle C_{j,n_{\mathrm{ph}}}\rangle$ contains the phase information needed to evaluate the dispersion relation of the $n_{\mathrm{ph}}$ bound states: 
\begin{multline}
    \ket{\psi(t)}= \frac{1}{\sqrt{2}} \biggl( \ket{0}^{\otimes N_Q}+\sum_k \alpha_k e^{-i\omega(k)t}\ket{k}\biggr ) \\
    \rightarrow \, \langle C_{j,n_{\mathrm{ph}}}\rangle = 1/(2\sqrt{N_Q}) \sum_{k} \alpha^*_k e^{i(\omega(k) t-kj)}, 
\end{multline}
\noindent
where $\ket{k}$ and $\alpha_k$ are bound $n_{\mathrm{ph}}$-photon momentum states and their coefficients, respectively. 

\vspace{2mm}  
Fig.\,\ref{fig:fig3}b shows the real and imaginary parts of the correlator for the case of two photons. While the real space data displays a rather intricate pattern (Fig.\,\ref{fig:fig3}b), conversion to the energy and momentum domain through a 2D Fourier transform reveals a clear band structure for both the single-particle and the many-body states (Fig.\,\ref{fig:fig3}c). The observed bands, which are defined modulo $2\pi$/cycle due to the discrete time translation symmetry of the Floquet circuit, are in agreement with the predictions of Eq. \ref{eq:dispersion_relation}, as illustrated in colored dashed curves. The bands shift when the photon number increases, as expected from the higher total interaction energy. Moreover, they become flatter, a characteristic feature of increased interaction effects.

In order for a bound state to form, the interaction energy must be sufficiently high compared to the kinetic energy of the particles. In particular, bound states are only expected to exist for all momenta when $\phi>2\theta$\,\cite{Aleiner2021}. To explore this dependence on $\phi/\theta$, we also measure the band structure for $n_{\mathrm{ph}}=2$ in the weakly interacting regime ($\theta=\pi/3$, $\phi=\pi/6$; Fig.\ref{fig:fig3}d), which exhibits very different behavior from the more strongly interacting case studied in Fig.\ref{fig:fig3}c: while no band is observed for most momenta, a clear state emerges near $k=\pm\pi/\mathrm{site}$. Interestingly, this observation of a bound state in the weakly interacting regime can be attributed to destructive interference of the decay products of the bound state: a 2-photon bound state $\ket{..0110..}$ can separate into two possible states, $\ket{..1010..}$ and $\ket{..0101..}$, which are shifted relative to each other by one lattice site. Hence, they destructively interfere when the momentum is near $k=\pm\pi/\mathrm{site}$, which prevents separation. (See the supplementary information for band structures of additional fSim angles.)     

\begin{figure*}
    \centering
    \includegraphics[width=\textwidth]{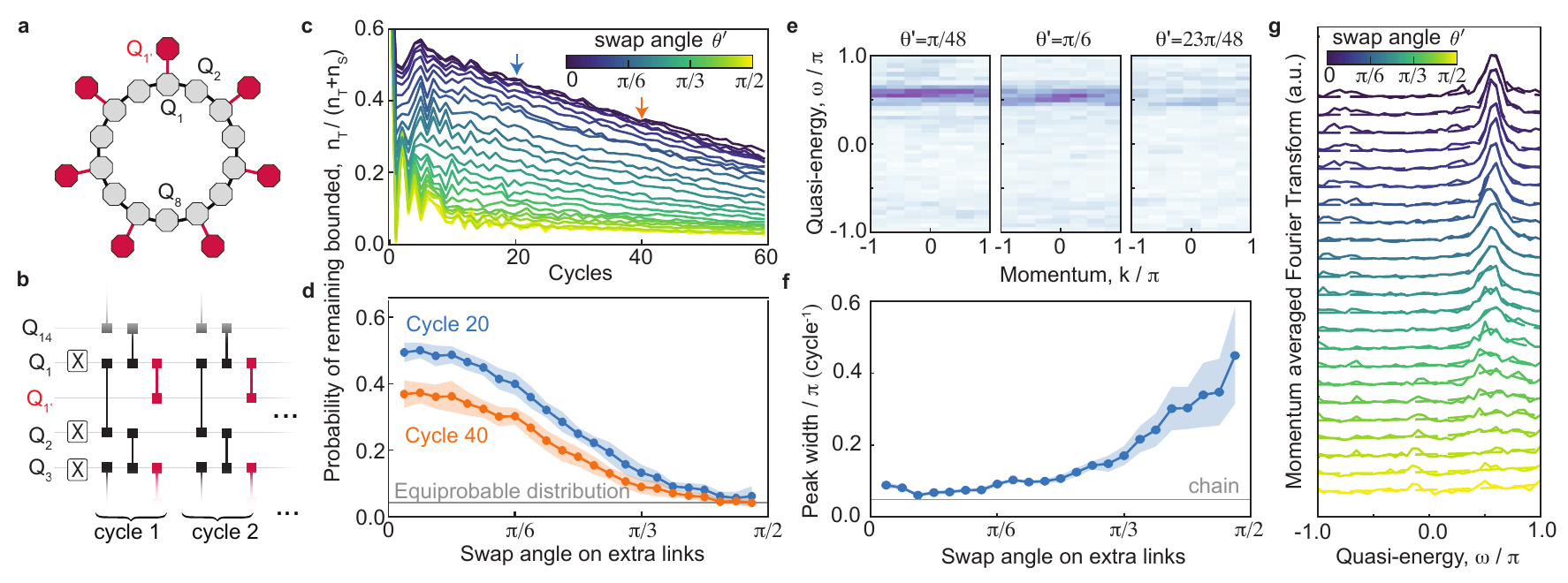}
    \caption{\textbf{Resilience to integrability breaking. a}, Schematic of the 14 qubits chain with 7 extra sites in red to break the integrability. \textbf{b}, Integrability is broken via an extra layer of fSim-gates (red) between the chain and the extra qubits, with $\phi'=\phi$ and a gradually varied $\theta'$. \textbf{c}, Decaying probability of remaining bound for different swap angles $\theta'$. Similar to Fig.\,\ref{fig:fig2}e, the bound state decays into the continuum due to the dephasing. \textbf{d}, Probability of remaining bound after 20 and 40 cycles as $\theta'$ is swept. \textbf{e}, Spectroscopy of the $n_{\mathrm{ph}}=3$ bound state for different $\theta'$. Note that the bound state survives even for $\theta'=\theta$. \textbf{f}, Half width of the momentum averaged spectra (\textbf{g}) as a function of $\theta'$. The gray line indicates the result for the chain without the extra qubits. \textbf{g}, Momentum averaged quasi-energy spectra for varying $\theta'$ fitted with Lorentzian. The bound state peak slowly disappears with the increase of $\theta'$.}
    \label{fig:fig4}
\end{figure*}

External magnetic fields can shift the energy bands and reveal the electric pseudo-charge of the quasi-particles constituting the band. We produce a synthetic magnetic flux $\Phi$ that threads the ring of qubits by performing $Z$-rotations with angles $\pm\Phi/N_Q$ on the qubits before and after the two-qubit fSim gates, resulting in a complex hopping phase $\beta=\Phi/N_Q$ when a photon moves from site $j$ to $j+1$\,\cite{Charles2021}. As a consequence, the eigenstates are expected to attain a phase $(n_{\mathrm{ph}}\beta)\cdot j$, effectively shifting their quasi-momentum by $n_{\mathrm{ph}}\beta$. Fig.\,\ref{fig:fig3}e displays the flux dependence of the two-photon band structure, exhibiting a clear shift in momentum as $\Phi$ increases. In Fig.\,\ref{fig:fig3}f, we extract the shift for $n_{\mathrm{ph}}=1-5$ and observe excellent agreement with the theoretical predictions\,\cite{Aleiner2021}. Crucially, the momentum shift is found to scale linearly with $n_{\mathrm{ph}}$, indicating that the observed states have the correct pseudo-charge. 

Generally, bound states in the continuum are rare and very fragile, and their stability rely on integrability or symmetries \cite{BIC_continuum, Groha_2017}.
Familiar stable dimers, such as excitons in semiconductors, have energy resonances in the spectral gap. In the system considered here, the bound states are predicted to almost always be inside the continuum due to the periodicity of the quasi-energy. Our results shown in Figs.\,\ref{fig:fig3} demonstrate an experimental verification of this remarkable theoretical prediction in the integrable limit and constitutes our first major result. 

Next we probe the stability of the bound states against integrability breaking. Fermi's golden rule suggests that any weak perturbation that breaks the underlying symmetry will lead to an instability and a rapid decay of the bound states into the continuum. We examine the robustness of the $n_{\mathrm{ph}}=3$ bound state by constructing a quasi-1D lattice where every other site of the 14 qubit ring is coupled to an extra qubit site\,(Fig.\,\ref{fig:fig4}a). The extra sites increase the Hilbert space dimension and ensure that the system is not integrable. We implement the circuit depicted in Fig.\,4b with three layers of fSim gates in each cycle. The first two layers are the XXZ ring dynamics with the same parameters used in Fig.\,2: $\theta=\pi/6$ and $\phi=2\pi/3$. In the third layer we also use $\phi'=2\pi/3$ but vary the swap angle $\theta'$ to tune the strength of the integrability breaking perturbation.  

Fig.\,\ref{fig:fig4}c shows the probability of measuring three-photon $\mathcal{T}$-bitstrings as a function of time for various $\theta'$ angles. In the limit of small $\theta'$, where the integrability breaking is weak, the system shows a slowly decaying probability, similar to the unperturbed (integrable, $\theta'=0$) results presented in Fig\,2. In Fig.\,\ref{fig:fig4}d, we show the dependence of this probability on perturbation strength after two fixed circuit depths. For strong perturbations, the integrability breaking washes out the bound state and the probability rapidly decays to the equiprobable distribution in the full Hilbert space of 3 photons ($\mathcal{T}$+$\mathcal{S}$). However, the surprising finding is that even up to $\theta' = \pi/6$, which corresponds to perturbation gates identical to the gates on the main ring, \textit{i.e.} a strong perturbation, there is very little decay in $n_{\mathcal{T}}$. This observation demonstrates the resilience of the bound state to perturbations far beyond weak integrability breaking for $n_{\mathrm{ph}}=3$. We further confirm this finding by performing spectroscopy of these states, which shows the presence of the $n_{\mathrm{ph}}=3$ bound states up to large perturbations (Fig.\,\ref{fig:fig4}e). By fitting the momentum averaged spectra (Fig.\,\ref{fig:fig4}g), we extract the $\theta'$-dependence of the half-width of the band (Fig.\,\ref{fig:fig4}f). Indeed, we find that the bandwidth is insensitive to $\theta'$ up to very large perturbation. 

These observations are at odds with the expectation that non-integrable perturbation leads to the fast decay of bound states into the continuum. One known exception is many-body scars, where certain initial states exhibit periodic revivals and do not thermalize\,\cite{Bluvstein2021,Turner2017}. Moreover, in the case of weak integrability breaking, robustness to perturbations can result from quasi-conserved or hidden conserved quantities\,\cite{Brandino2015, Kurlov2022}. However, the resilience observed here extends well beyond the weak integrability breaking regime typically considered in such scenarios. Alternatively, the presence of highly incommensurate energy scales in the problem can lead to a very slow decay in a chaotic system due to parametrically small transition matrix elements, a phenomenon called prethermalization\,\cite{Strohmaier2010, Abanin_2017}. Our experiment finds the survival of an integrable system's feature  - bound states - for large perturbation and in the absence of obvious scale separation, which may point to a new regime arising due to interplay of integrability and prethermalization.

The key enabler of our experiment is the capability of tuning high fidelity $\FSIM$ gates to change the ratio of kinetic to interaction energy, as well as directly measuring multi-body correlators $\langle C_{j,n_{\mathrm{ph}}}\rangle$, both of which are hard to access in conventional solid state and atomic physics experiments. Aided by these capabilities, we observed the formation of multi-photon bound states and discovered a striking resilience to non-integrable perturbations. This experimental finding, although still observed for computationally tractable scales, in the absence of any theoretical prediction, constitutes our second major result\,(Fig.\,4). A proper understanding of this unexpected discovery is currently lacking.   

\newpage
\onecolumngrid
\vspace{1em}
\begin{flushleft}
    {\hypertarget{authorlist}{${}^\dagger$}  \small Google Quantum AI and Collaborators}

    \bigskip

    \renewcommand{\author}[2]{#1\textsuperscript{\textrm{\scriptsize #2}}}
    \renewcommand{\affiliation}[2]{\textsuperscript{\textrm{\scriptsize #1} #2} \\}
    \newcommand{\corrauthora}[2]{#1$^{\textrm{\scriptsize #2}, \hyperlink{corra}{\ddagger}}$}
    \newcommand{\corrauthorb}[2]{#1$^{\textrm{\scriptsize #2}, \hyperlink{corrb}{\mathsection}}$}
    \newcommand{\xGoogle}{\affiliation{1}{Google Research, Mountain View, CA, USA}}

\newcommand{\xGeneva}{\affiliation{2}{Department of Theoretical Physics, University of Geneva, Quai Ernest-Ansermet 30, 1205 Geneva, Switzerland}}

\newcommand{\xUMass}{\affiliation{3}{Department of Electrical and Computer Engineering, University of Massachusetts, Amherst, MA, USA}}

\newcommand{\xYale}{\affiliation{4}{Department of Applied Physics, Yale University, New Haven, CT 06520, USA}}

\newcommand{\xCaltech}{\affiliation{5}{Institute for Quantum Information and Matter, California Institute of Technology, Pasadena, CA, USA}}

\newcommand{\xUCR}{\affiliation{6}{Department of Electrical and Computer Engineering, University of California, Riverside, CA, USA}}

\newcommand{\xUCSB}{\affiliation{7}{Department of Physics, University of California, Santa Barbara, CA, USA}}

\begin{footnotesize}

\newcommand{\Google}{1}
\newcommand{\Geneva}{2}
\newcommand{\UMass}{3}
\newcommand{\Yale}{4}
\newcommand{\Caltech}{5}
\newcommand{\UCR}{6}
\newcommand{\UCSB}{7}

\corrauthora{A. Morvan}{\Google},
\corrauthora{T. I.~Andersen}{\Google},
\corrauthora{X. Mi}{\Google},
\corrauthora{C. Neill}{\Google},
\author{A. Petukhov}{\Google},
\author{K. Kechedzhi}{\Google},
\author{D. A. Abanin}{\Google,\! \Geneva},
\author{R. Acharya}{\Google},
\author{F. Arute}{\Google},
\author{K. Arya}{\Google},
\author{A. Asfaw}{\Google},
\author{J. Atalaya}{\Google},
\author{R. Babbush}{\Google},
\author{D. Bacon}{\Google},
% ...
\corrauthorb{I. Aleiner}{\Google},
\corrauthorb{L. B.~Ioffe}{\Google},
\corrauthorb{P. Roushan}{\Google}

\bigskip

\xGoogle
\xGeneva
\xUMass
\xYale
\xCaltech
\xUCR
\xUCSB

{\hypertarget{corra}{${}^\ddagger$} These authors contributed equally to this work.}\\
{\hypertarget{corrb}{${}^\mathsection$} Corresponding author: igoraleiner@google.com}\\
{\hypertarget{corrb}{${}^\mathsection$} Corresponding author: ioffel@google.com}\\
{\hypertarget{corrb}{${}^\mathsection$} Corresponding author: pedramr@google.com}
\end{footnotesize}

\end{flushleft}

\twocolumngrid

\bibliography{main.bib}
\end{document}


\title{Supplement - Formation of robust bound states of interacting photons}
\author{Google Quantum AI and Collaborators}
\date{\today}

\maketitle
\tableofcontents

\newpage

\section{Quantum processor details and coherence times}\label{gatecalib}

\begin{figure}[h]
    \centering
    \includegraphics[width=0.6667\columnwidth]{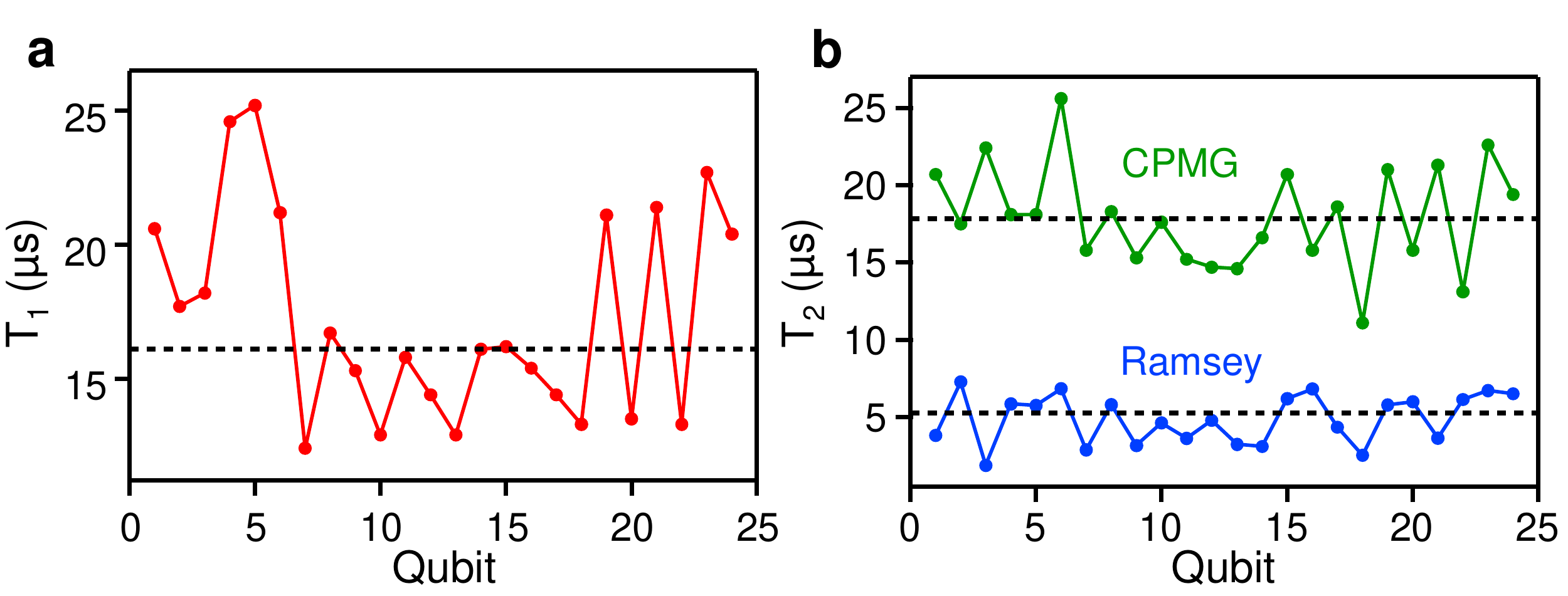}
    \caption{\textbf{a,} Single-qubit relaxation times $T_1$ across the 24-qubit ring. Dashed line represents the median value of 16.1 $\upmu$s. \textbf{b,} Single-qubit $T_2$ values measured via Ramsey and Carr-Purcell-Meiboom-Gill (CPMG) dynamical decoupling sequence. Dashed lines indicate median $T_2$ values of 5.3 $\upmu$s (Ramsey) and 17.8 $\upmu$s (CPMG), respectively.}
    \label{fig:coherence}
\end{figure}

The experiment is performed on a quantum processor with similar design as that in Ref.~\cite{Arute2019}. The qubits are transmons with tunable frequencies and interqubit couplings. Figure~\ref{fig:coherence}a shows the single-qubit relaxation times of the 24 qubits used in the experiment, where a median value of $T_1 = 16.1$ $\upmu$s is found. The dephasing times $T_2^*$, measured via Ramsey interferometry, are shown in Fig.~\ref{fig:coherence}b and have a median value of 5.3 $\upmu$s. Lastly, The $T_2$ values after CPMG dynamical decoupling sequences are also shown in Fig.~\ref{fig:coherence}b and have a median of 17.8 $\upmu$s.

\section{2-Qubit fSim gates}\label{sec2}

\subsection{fSim calibration }\label{fsim_calibration}
\begin{figure}[h]
    \centering
    \includegraphics[width=0.6667\columnwidth]{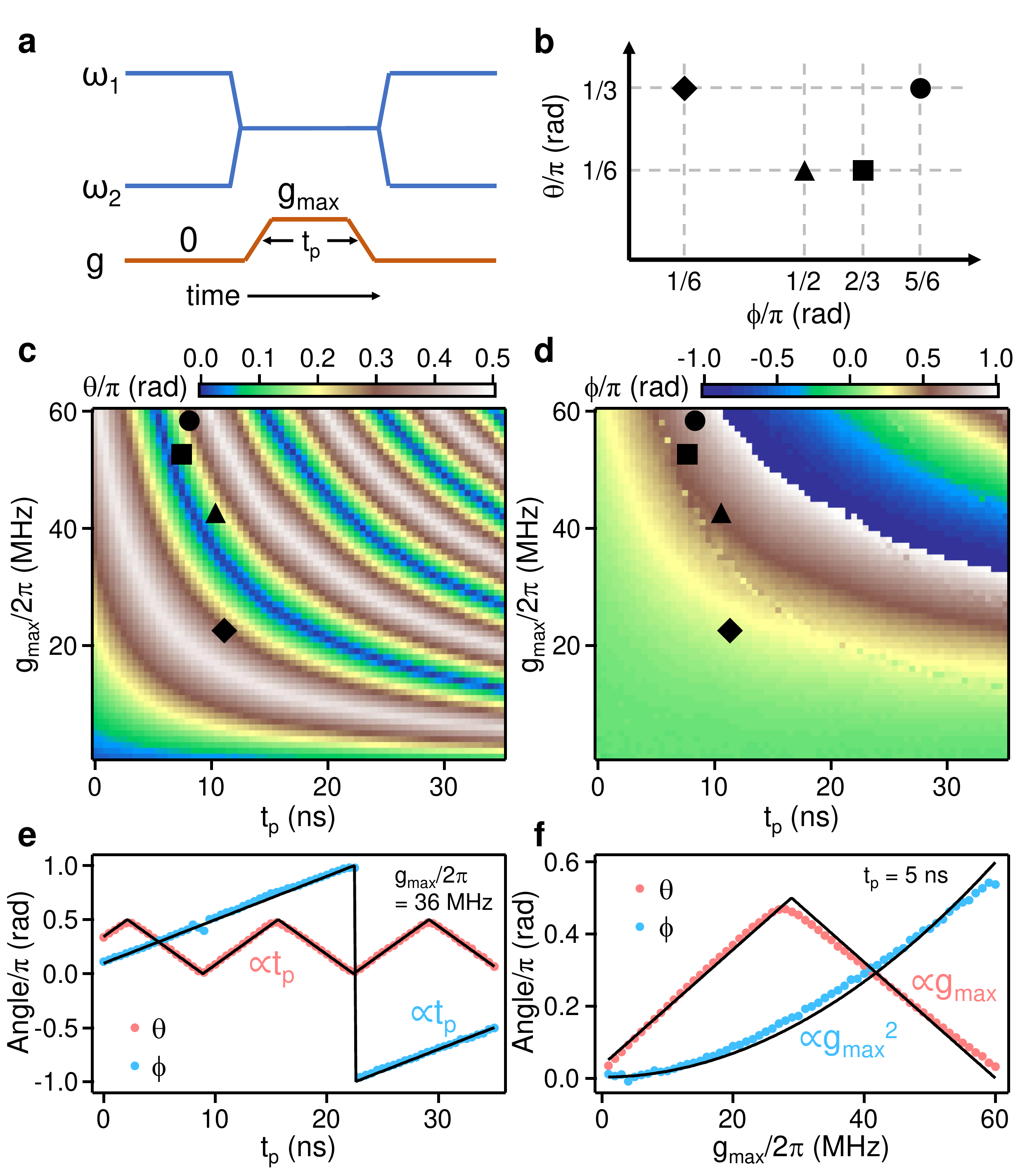}
    \caption{\textbf{a,} Schematic illustration of the DC pulse used to realize a fSim gate. \textbf{b,} Schematic plot showing the four sets of fSim angles used in this work: $(\theta / \pi, \phi / \pi) = (\frac{1}{3}, \frac{5}{6})$ (filled circle), $(\frac{1}{6}, \frac{2}{3})$ (square), $(\frac{1}{6}, \frac{1}{2})$ (triangle) and $(\frac{1}{3}, \frac{1}{6})$ (diamond). \textbf{c,} Experimentally measured $\theta$ as a function of pulse length $t_{\text p}$ and maximum interqubit coupling $g_{\text max}$. The approximate pulse parameters for the fSim gates in panel b are indicated by their corresponding symbols. \textbf{d,} Similar plot as panel c but with $\phi$ shown instead. \textbf{e,} $\theta$ and $\phi$ as functions of $t_{\text p}$ for a fixed $g_{\text max} / {2 \pi}$ of 36 MHz. Solid lines show linear fits. \textbf{f,} $\theta$ and $\phi$ as functions of $g_{\text max}$ for a fixed $t_{\text p}$ of 5 ns. Solid lines show a linear fit $\propto g_{\text max}$ to $\theta (g_{\text max})$ and a quadratic fit $\propto g_{\text max}^2$ to  $\phi (g_{\text max})$.}
    \label{fig:fsim_scan}
\end{figure}

The floquet unitaries used in the experiment are composed of alternating layers of $\FSIM(\theta, \phi,\beta)$ gates which are defined as:
\begin{equation}
    \FSIM(\theta, \phi, \beta) = \begin{pmatrix} 1 & 0 & 0 & 0 \\
                                         0 & \cos \theta & ie^{i \beta} \sin \theta & 0 \\
                                         0 & i e^{-i\beta}\sin \theta & \cos \theta & 0 \\
                                         0 & 0 & 0 & e^{i\phi} \\
    \end{pmatrix},
\end{equation}
\noindent
where $\theta$ is the SWAP angle and $\phi$ is the conditional phase, and $\beta$ is phase accumulated resulting from hopping. For open chains, $\beta$ is not gauge invariant and can be ignored, but for closed chains, non-zero $\beta$ values lead to a total flux threading the closed chain. $\FSIM(\theta, \phi, \beta)$ describes the unitary form output by a DC pulse bringing the fundamental frequencies $\omega_1$ and $\omega_2$ of two transmons into resonance and turning on their interqubit coupling $g$ for a given time duration $t_{\text p}$, as illustrated in Fig.~\ref{fig:fsim_scan}a. During $t_{\text p}$, resonant interaction between the $\ket{10}$ and $\ket{01}$ states of the two transmons leads to population transfer and a finite $\theta$. Additionally, dispersive interaction between the $\ket{11}$ and $\ket{02}$ (as well as $\ket{20}$) states of the two qubits gives rise to a finite conditional phase $\phi$.

Due to the frequency detunings of the qubits during the DC pulse, the fSim unitary also includes additional single-qubit $Z$ rotations and is more generally described by:
\begin{equation}\label{fsim_def}
    \text{FSIM}(\theta, \phi, \gamma, \alpha, \beta) = \begin{pmatrix} 1 & 0 & 0 & 0 \\
                                         0 & e^{i(\gamma - \alpha)} \cos \theta  & i e^{i(\gamma + \beta)} \sin \theta  & 0 \\
                                         0 & i e^{i(\gamma - \beta)} \sin \theta  & e^{i(\gamma + \alpha)} \cos \theta & 0 \\
                                         0 & 0 & 0 & e^{2i\gamma}e^{i\phi} \\
    \end{pmatrix},
\end{equation}
where $\gamma$, $\alpha$ and $\beta$ are complex phases incurred by the single-qubit $Z$ rotations. These single-qubit phases are calibrated and reduced to nearly zero using the technique of Floquet calibration described in our previous works \cite{Charles2021,zhang_floquet_2020,DTC_Nature_2022}. Here we focus on the tuning and calibration of the two-qubit angles $\theta$ and $\phi$.

Figure~\ref{fig:fsim_scan}c and Figure~\ref{fig:fsim_scan}d show experimentally obtained values of $\theta$ and $\phi$ as functions of pulse parameters $t_{\text p}$ and $g_{\text max}$. In these measurements, the technique of unitary tomography \cite{Foxen_PRL_2020} is used to estimate the angles. We have also enforced a Gaussian filter with time constant 5 ns on the rising/falling edges of the pulse on $g$ to ensure adiabatic evolution with respect to the $\ket{11} \rightarrow \ket{20}$ and $\ket{11} \rightarrow \ket{02}$ transitions. This is important to minimize leakage. We observe that $\theta$ shows a series of maxima/minima corresponding to values of $t_{\text p}$ and $g_{\text max}$ where $\ket{01}$ is fully transformed to $\ket{10}$ or returned back to $\ket{01}$. On the other hand, $\phi$ increases monotonically over $t_{\text p}$ and $g_{\text max}$ until it reaches a maximum value of $\pi$ where it is wrapped by $2 \pi$ and becomes $-\pi$. Given the dependence of both unitary angles on both pulse parameters $t_{\text p}$ and $g_{\text max}$, independent control of $\theta$ or $\phi$ is not possible with a single pulse parameter. As such, past works have added a resonant pulse between the $\ket{11}$ and $\ket{02}$ states to enact a pure CPHASE gate, thereby enabling full tunablity over $\theta$ and $\phi$ \cite{Foxen_PRL_2020}. The additional pulse, however, significantly increases the complexity of quantum control and is also prone to leakage. 

In this work, we have chosen to perform fSim gates directly using the single pulse in Fig.~\ref{fig:fsim_scan}a. Our approach relies on the different scaling of $\theta$ and $\phi$ with the pulse parameters, as illustrated by Fig.~\ref{fig:fsim_scan}e and Fig.~\ref{fig:fsim_scan}f. Here we observe that while $\theta$ and $\phi$ both scale linearly with $t_{\text p}$, the scaling with $g_{\text max}$ is different for the two angles: whereas $\theta$ scales linearly with $g_{\text max}$, $\phi \propto g_{\text max}^2$ due to the fact that dispersive shift of the $\ket{11}$ state by the $\ket{02}$ and $\ket{20}$ states is proportional to $g^2 / \Delta$, where $\Delta$ is the frequency difference between $\ket{11}$ and the $\ket{02}$ ($\ket{20}$). The difference in scaling implies that it is possible to achieve a desired combination of $\theta$ and $\phi$ by choosing a particular ``contour'' in the 2D space of ($t_{\text p}$, $g_{\text max}$) where $\theta$ has the target value, then increasing (decreasing) $g_{\text max}$ while decreasing (increasing) $t_{\text p}$ until $\phi$ attains the target value as well.

Practically, $\phi$ and $\theta$ are calibrated via a simple gradient descent method: we start with an initial guess ($t_0$, $g_0$) for the pulse parameters ($t_{\text p}$, $g_{\text max}$) based on the 2D scan shown in Fig.~\ref{fig:fsim_scan}c and Fig.~\ref{fig:fsim_scan}d. The corresponding values of $\phi$ and $\theta$ are then accurately determined via Floquet calibration \cite{zhang_floquet_2020,Charles2021, DTC_Nature_2022} which we denote as $\phi_0$ and $\theta_0$. We then calibrate the fSim angles at $(t_{\text p}, g_{\text max}) = (t_0 + \delta t, g_0)$ and $(t_{\text p}, g_{\text max}) = (t_0, g_0 + \delta g)$. The results allow us to approximate the following gradient matrix:
\begin{equation}
M_{\text g} = 
\begin{pmatrix}
{\partial \phi}/{\partial t} \,\,& {\partial \phi}/{\partial g} \\ 
{\partial \theta}/{\partial t} & {\partial \theta}/{\partial g}
\end{pmatrix}.
\end{equation}
A new set of pulse parameters $(t_1, g_1)$ are then computed from the gradient matrix and the deviations from target fSim angles, $(\Delta t, \Delta g) = (t_{\text c} - t_0, g_{\text c} - g_0)$, via:
\begin{equation}
\begin{pmatrix}
t_1  \\
p_1
\end{pmatrix} = 
M_{\text g}^{-1} \begin{pmatrix}
\Delta t  \\
\Delta g
\end{pmatrix} + \begin{pmatrix}
t_0  \\
p_0
\end{pmatrix}.
\end{equation}
The fSim angles $(t_{\text p}, g_{\text max})$ are then measured at the new pulse parameters $(t_1, g_1)$ and the process is repeated. Generally only two gradient descents are sufficient to reach control errors on the level of $\sim$20 mrad for both $\theta$ and $\phi$.

When repeating the fSim gate $n$ times in Floquet calibration, the error accumulates as:
\begin{equation}\label{fSimDef}
    \FSIM(\theta, \phi) = \begin{pmatrix} 1 & 0 & 0 & 0 \\
                                         0 &  e^{i\gamma n}A & ie^{i(\gamma n + \beta)} B  & 0 \\
                                         0 & i e^{i(\gamma n - \beta)} B & e^{i\gamma n} A^{*} & 0 \\
                                         0 & 0 & 0 & e^{2i\gamma n}e^{i\phi n} \\
    \end{pmatrix}
\end{equation}
with \begin{align}
    A & = \cos n\Omega - iQ \sin n\Omega \\
    B & = \sqrt{1-Q^2} \sin n \Omega \\
    \cos \Omega & = \cos \theta \cos \alpha/2 \\
    Q & = \cos \theta \sin \alpha/2 /\sin \Omega
\end{align}

\begin{figure}
    \centering
    \includegraphics{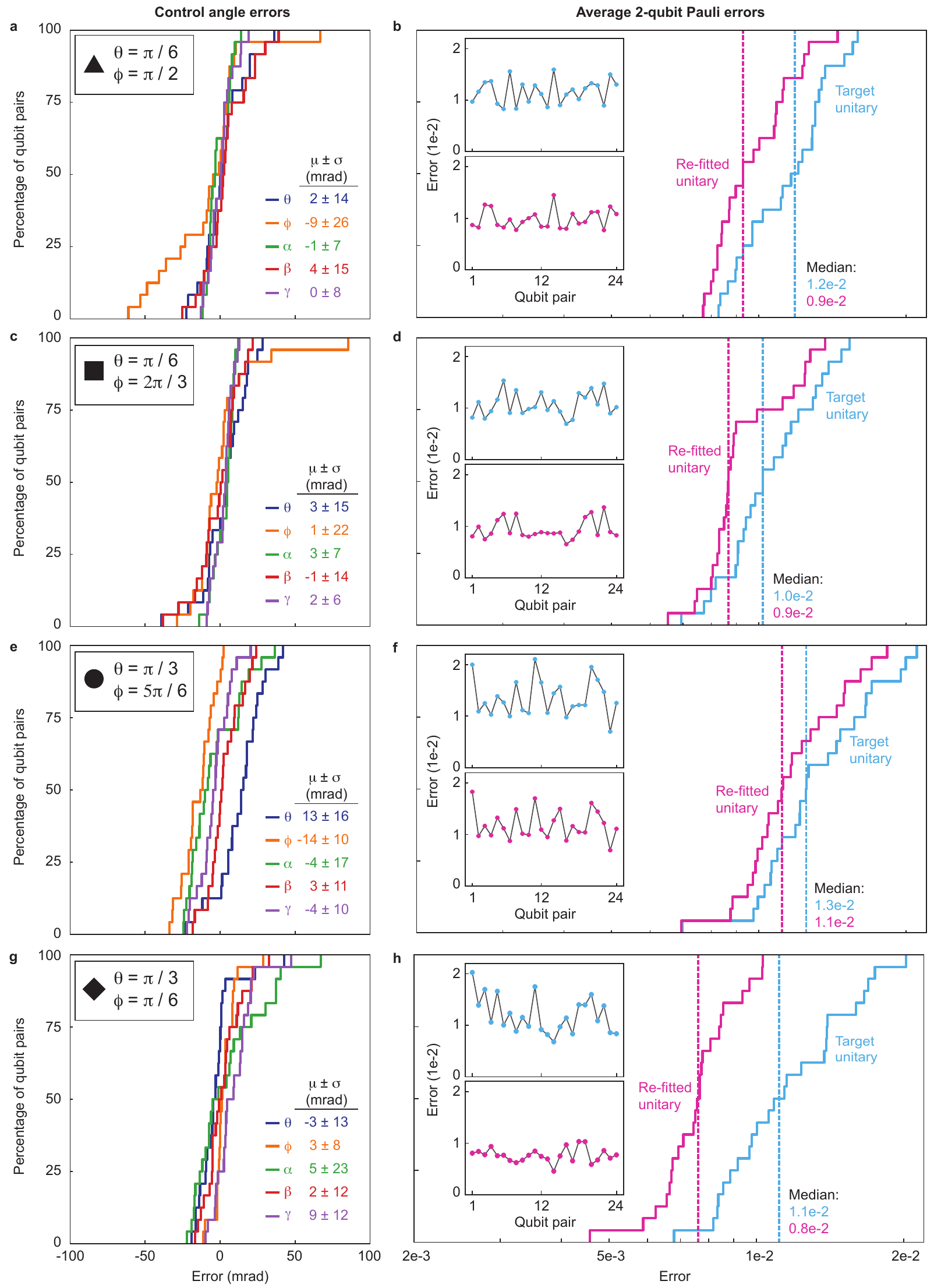}\label{fig:ControlErrorFig}
    \caption{\textbf{a}, Integrated histogram of errors in the control angle parameters of the fSim gate as measured by periodic calibration, for target angles $\theta=\pi/6$, $\phi=\pi/2$, $\alpha=\beta=\gamma=0$. Inset: table displaying the mean errors and their standard deviation. \textbf{b}, Integrated histogram of average 2-qubit Pauli errors, considering both the target unitary (blue) and the re-fitted unitary (pink). Inset: Pauli errors plotted against qubit position, demonstrating the absence of systematic spatial dependence. Dashed vertical lines denote the medians. \textbf{c-h}, Same as \textbf{a,b}, but for target fSim angles, $\theta=\pi/6$, $\phi=2\pi/3$ (\textbf{c,d}), $\theta=\pi/3$, $\phi=5\pi/6$ (\textbf{e,f}), $\theta=\pi/3$, $\phi=\pi/6$ (\textbf{g,h}). The remaining target angles are zero ($\alpha=\beta=\gamma=0$) in all plots.}
\end{figure}

\subsection{fSim gate control error}
The angular errors, which are measured using periodic Floquet calibration as outlined in section \ref{fsim_calibration}, are displayed in Fig.\ref{fig:ControlErrorFig} along with the measured 2-qubit Pauli error for the four pairs of $\theta$,$\phi$ studied in our work. The angular errors of the fSim gate can be combined into an overall control error by calculating the coherent gate infidelity:
\begin{align}
\epsilon_c=1-F=\frac{1}{1+D^{-1}}\overbrace{\left(1-\frac{1}{D^2}|\mathrm{Tr}\{U_{\mathrm{target}}^{\dagger}U_{\mathrm{actual}}\}|^2\right)}^{e_{\mathrm{P}}}
\end{align}
where $e_{\mathrm{P}}$ is the Pauli error, $D=4$ is the size of the computational subspace of the gate, and $U_{\mathrm{target(actual)}}$ is the target (actual) unitary. Inserting the unitary matrix in eq. \ref{fsim_def}, we find the control error in terms of the angular errors:
\begin{align}
\epsilon_c=\frac{2}{5}d\theta^2+\frac{3}{20}d\phi^2+\frac{1}{10}d\gamma^2+\frac{1}{5}d\gamma d\phi+\frac{1}{10}\cos^2\theta\cdotp d\alpha^2+\frac{1}{80}\sin^2\theta\left(7+\cos2\theta+2\sin^2\theta\right)\cdotp d\beta^2
\end{align}
For the four angle pairs in our study, we find median control errors of $\epsilon_c(\theta=\pi/6, \phi=2\pi/3)=1.5\cdotp10^{-4}$, $\epsilon_c(\theta=\pi/6, \phi=\pi/2)=1.0\cdotp10^{-4}$, $\epsilon_c(\theta=\pi/3, \phi=5\pi/6)=3.1\cdotp10^{-4}$, and $\epsilon_c(\theta=\pi/3, \phi=\pi/6)=1.8\cdotp10^{-4}$.

\newpage

\section{Analytical results for the Floquet XXZ chain}

\subsection{Band structure}
In this section, we present analytical expressions for the dispersion relations in the Floquet XXZ model, as derived in Ref.\cite{aleiner2021}. The solution is separated into two cases, corresponding - in the limit of the Hamiltonian model - to the gapped ($\phi > 2\theta$) and the gapless ($\phi < 2\theta$) phase. 

We first focus on the gapped phase ($\phi > 2\theta$), since in this case, there exists a bound state for all $n_{\mathrm{ph}}$ for every momentum (see end of section 3.5 p.19 in Ref.\cite{aleiner2021}). Importantly, the dispersion relations of the bound states have a similar functional form to that of a single particle excitation (see Eqs. (72) and (85) in \cite{aleiner2021}):
\begin{equation}
    \cos(E(k) - \chi) = \cos^2(\alpha) - \sin^2(\alpha) \cos(k)  \quad \text{with} \quad k \in [-\pi, \pi],
    \label{eq_s1}
\end{equation}
where $\alpha$ and $\chi$ depend on the number of photons:
\begin{itemize}
    \item For $n_{\mathrm{ph}} = 1$, the parameters are $\alpha=\theta$ and $\chi=0$
    \item For $n_{\mathrm{ph}} \geq 2$, one finds:
        \begin{equation}
            \chi = n_{\mathrm{ph}}\phi - 2 \arctan\left(\tan (\phi/2) \tanh \eta \coth(n_{\mathrm{ph}}\eta)\right)
        \end{equation}
        \begin{equation}
            \cos^2\alpha = \frac{\cos^2\theta \sinh^2 n_{\mathrm{ph}}\eta}{\cos^2\theta\sinh^2 n_{\mathrm{ph}}\eta+\sin^2 \theta \sinh^2 \eta},
        \end{equation}
        where $\eta$ is given by:
        \begin{equation}
            \sinh^2(\eta) = \frac{\cos^2\theta - \cos^2(\phi/2)}{\sin^2\theta}.
            \label{eq_s2}
        \end{equation}
\end{itemize}

In the gapless regime $(\phi<2\theta)$, the bound state is only predicted to appear for one of the two branches in the first Brillouin zone and for a finite range of momenta, with limits given by (see end of section 3.5 p.19 in Ref.\cite{aleiner2021}): 
\begin{equation}
k_0=\pm2n_{\mathrm{ph}}\eta
\end{equation}
These are the limits plotted with dashed black vertical lines in Fig. 3d in the main text, as well as in Fig. \ref{fig:bandstructures}d. In this case the dispersion relation is similar to the gapless case with the substitution $\sinh \eta \mapsto \sin \eta$ and $\tanh \eta \mapsto \tan \eta$.

\subsection{Group velocity}

We can calculate the group velocity by differentiating both sides of \ref{eq_s1}:
\begin{equation}
    v_{g}(k) = \frac{\mathrm{d} E}{\mathrm{d}k}=  -\frac{\sin^2 \alpha \sin k}{\sin( E(k) - \chi )}
\end{equation}
We notice that the maximum group velocity is achieved for $k=\pi/2$ leading to
\begin{equation}
    v_g^{\text{max}}= \frac{\sin^2 \alpha}{\sqrt{1-\cos^4 \alpha}}
\end{equation}
This is the velocity of the initial cone that is shown in the main text Fig.2d

\newpage
\subsection{Choice of parameters \texorpdfstring{$\theta$}{TEXT} and \texorpdfstring{$\phi$}{TEXT}}
The choice of the angles $\theta$ and $\phi$ of the fSim gate is dictated by several considerations, with regards to both the physics of the bound state and the experimental parameters of the fSim gate. In the main text, we present several angles as they have differents properties:

\begin{figure}
    \centering
    \includegraphics{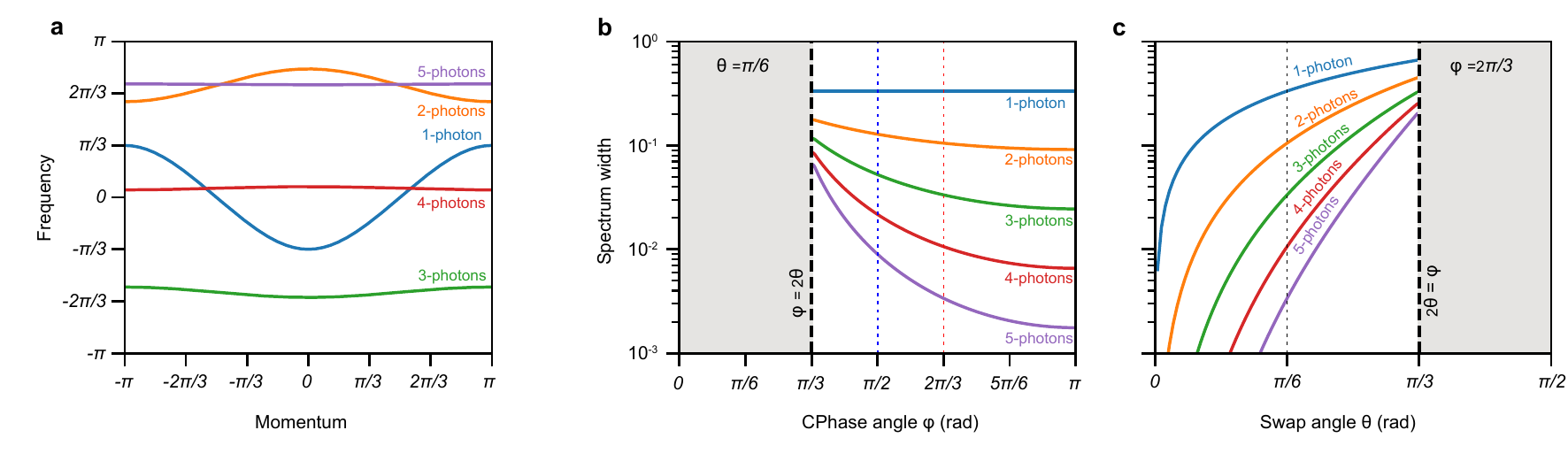}
    \caption{\textbf{a}, Band structure of the different bound states for $\theta = \pi/6$ and $\phi=2\pi/3$. \textbf{b, c}, Width of the $n_{\mathrm{ph}}$-photon band structure as a function of the parameters of the \FSIM~gate. All these results are calculated using the exact solution given in eq. \ref{eq_s1}.}
    \label{fig:supp_spectrum_width}
\end{figure}

\begin{itemize}
    \item First and foremost, the behavior is significantly different for the gapped $\phi > 2\theta$ regime and the gapless regime. We have predominantly focused on the gapped regime where bound states exist for all momenta, simplifying the analysis.
    \item Secondly, as shown in Fig. \ref{fig:supp_spectrum_width}b, the amount of dispersion in the spectrum increases with increasing (decreasing) values of $\theta$ ($\phi$). Hence, the ratio $\phi/\theta$ should not be too large in order to maximize the visibility of the dispersion.
    \item Higher values of $\phi$ cause the bound state to be more localized, as intuitively expected from the stronger interactions. 
    \item Finally the quality of the fSim gate depends on the angles, since strong interaction requires longer gates time or larger interaction strength (see section \ref{sec2}).
\end{itemize}
\newpage

\section{Examples of multi-photon spectroscopy}

In this section we provide a pedagogical illustration of the basics of multi-photon spectroscopy by stepping through two simple examples. We focus on Hamiltonian dynamics, which is conceptually similar to the Floquet dynamics. The Fourier transform in space that was done in the main text is resulting from the space invariance of the circuit. For simplicity we only focus on the time-energy transformation and do not talk about the space-momentum transformation. 

\vspace{3mm}
Consider a Hamiltonian $\hat{\mathcal{H}}$ with eigenenergies $\omega_n$ and eigenstates $\ket{\varphi_n}$     
\begin{equation}
\hat{\mathcal{H}} \ket{\varphi_n} = \omega_n \ket{\varphi_n}.    
\end{equation}
\noindent
We seek a dynamical approach to learn the spectrum of $\mathcal{H}$. Consider an initial state  $\ket{\psi_0}$ and its evolution $\ket{\psi_t}$ \begin{equation}
\ket{\psi_0} = \sum_n  c_n \ket{\varphi_n},\,\,\,\,\,\,\,\,\,\,\,\,  \ket{\psi_t}= e^{-i\hat{\mathcal{H}}t}\ket{\psi_0} = \sum_n  c_n\,e^{-i\omega_nt}\ket{\varphi_n},
\end{equation}
\noindent where the complex coefficient $c_n=\bra{\varphi_{n}} \psi_{0} \rangle$ is the overlap of $\ket{\psi_0}$ with $\ket{\varphi_{n}}$. The desired spectrum of $\hat{\mathcal{H}}$ can be extracted (e.g. by Fourier transform) from the overlap of $\ket{\psi_0}$ and  $\ket{\psi_t}$
\begin{equation}
\bra{\psi_{0}} \psi_t \rangle = \sum_n  |c_n|^2 \, e^{-i\omega_nt}\,\,.
\end{equation}
\noindent
This overlap also frequently appears in non-equilibrium dynamics questions. Next, we provide a scheme to computing it via two simple examples. 

\vspace{3mm}

\textbf{Single-excitation example.} We begin with the case of single photon dynamics in a 4-qubit system. We are only considering excitation conserving Hamiltonians. The resulting unitary $\hat{U}$ when written in the basis of $\ket{0000}, \ket{1000},\ket{0100},\ket{0010}$ and $\ket{0001}$ becomes:

\begin{equation}
\begin{aligned}
\ket{0000} \,\, \ket{1000}  \,\, \ket{0100} \,\, \ket{0010} \,\,\, \ket{0001} \\
\hat{U}=\left(\begin{array}{ccccc} 
1 \,\,\,\, & \,\,\,\, 0 \,\,\,\,\,\,\, & 0       \,\,\,\,\,\, & \,\,\, 0 & \,\,\, 0\\ 
0 \,\,\,\, & \,\,\, u_{11} & \,\,\, u_{12} & \,\,\, u_{13}  & \,\,\, u_{14}\\
0 \,\,\,\, & \,\,\, u_{21} & \,\,\, u_{22} & \,\,\, u_{23} & \,\,\, u_{24}\\
0 \,\,\,\, & \,\,\, u_{31} & \,\,\, u_{32} & \,\,\, u_{33} & \,\,\, u_{34} \\
0 \,\,\,\, & \,\,\, u_{41} & \,\,\, u_{42} & \,\,\, u_{43} & \,\,\, u_{44} \\\end{array}\right). 
\end{aligned}
\end{equation}
\noindent
The first insight is using the fact that $\hat{U}$ is photon conserving and the $\ket{0000}$ manifold is detached from the other manifolds. We leverage this fact and create a phase-sensitive initial state (normalizations ignored) 

\begin{equation}
    \ket{\psi_0} =\ket{0000}+\ket{0100},
    \end{equation}
    \noindent which leads to 
    \begin{equation}
    \ket{\psi_t}= \hat{U}\ket{\psi_0}= \ket{0000}+u_{12}\ket{1000} +u_{22}\ket{0100} +u_{32}\ket{0010}+u_{42}\ket{0001}.
\end{equation}
\noindent From equations S19 to S21 it is evident that the spectrum is \textit{always} given by the Fourier transform of $\bra{\psi_{0}} \psi_t \rangle$. The task is to find an observable which is a measure of this overlap. Given the initial state we have chosen, then $\bra{\psi_{0}} \psi_t \rangle=1+u_{22}$. The second insight is to measure a non-hermitian operator $\hat{M}_{Q2} \equiv \ket{0000}\bra{0100}$, because:

\begin{equation}
\bra{\psi_t} \hat{M}_{Q2} \ket{\psi_t} = \bra{\psi_t} u_{22} \ket{0000} = u_{22}. \end{equation}

Note that the qubit lowering operator acting on Q2, $\sigma^-_2$, could also be written as
\begin{equation}
\sigma^-_2= \hat{X}_2 - i\hat{Y}_2 = \hat{M}_{Q2}\,\, 
\textcolor{gray} {+\ket{....}\bra{\text{more than single excitation}}}. 
\end{equation}
\noindent Since $\ket{\psi_t}$ has terms with only single photons, all gray colored terms can be ignored, and the Fourier transform of $\bra{\psi_t} \sigma^-_2 \ket{\psi_t}$ provides the desired spectrum. Notably, this would not have been achieved if the initial state was $\ket{\psi_0} = \ket{0100}$. While the overlap $\bra{\psi_{0}} \psi_t \rangle=u_{22}$ still gives the spectrum, we find $\bra{\psi_t} \hat{M}_{Q2} \ket{\psi_t}=0$ and also $\bra{\psi_t} \sigma^-_2 \ket{\psi_t}=0$, which clearly does not allow for measuring $u_{22}$.  

\vspace{5mm}
\textbf{Multi-excitation example.} Next, we consider the photon conserving dynamics of two excitations in a 4 qubit system, with the computational basis $\ket{1100}, \ket{1010},\ket{1001},\ket{0011},\ket{0101},\ket{00110}$. We start by placing Q3 and Q4 in a $\ket{0} + \ket{1}$ superposition, which gives rise to a superposition of the vacuum and a two-excitation state, as well as undesired single excitation states:
\begin{equation}
\ket{\psi_0} =\ket{0000}+\ket{0011}+\ket{0001} +\ket{0010}.   
\end{equation}
\noindent The appearance of single excitation terms is not desired and could have been avoided by using proper entangling gates to arrive at $\ket{0000}+\ket{0011}$ as a more relevant initial state. However, as our experimental results show, populating (wrong) manifolds with fewer number of excitations, is not harmful since the two-qubit lowering operator does not couple these manifolds to the vacuum. The existence of these undesired states does, however, reduce the signal contrast, due to distributed probabilities. The evolution of the initial states results in:    
\begin{multline}
\hat{U}\ket{\psi_0}= \ket{0000}+u_{14}\ket{1100} +u_{24}\ket{1010} +u_{34}\ket{1001}+u_{44}\ket{0011}+u_{54}\ket{0101}+u_{64}\ket{0110} \\
\textcolor{gray} {+\ket{\text{single excitation bitstrings}}}. 
\end{multline}
\noindent The two-qubit lowering operator acting on Q3 and Q4, can also be written as
\begin{equation}
    \sigma^-_3\sigma^-_4 = \hat{X}_3 \hat{X}_4-\hat{Y}_3 \hat{Y}_4+i\hat{X}_3 \hat{Y}_4+i\hat{Y}_3 \hat{X}_4=\ket{0000}\bra{0011}\textcolor{gray} {+\ket{....}\bra{\text{more than two excitations}}}.
\end{equation}
\noindent Hence, the presence of terms in other manifolds does not leading to wrong answers and $\bra{\psi_t} \sigma^-_3\sigma^-_4 \ket{\psi_t}$ provides the desired 2-photon spectrum.  
\begin{figure}
    \centering
    \includegraphics[width=0.6\columnwidth]{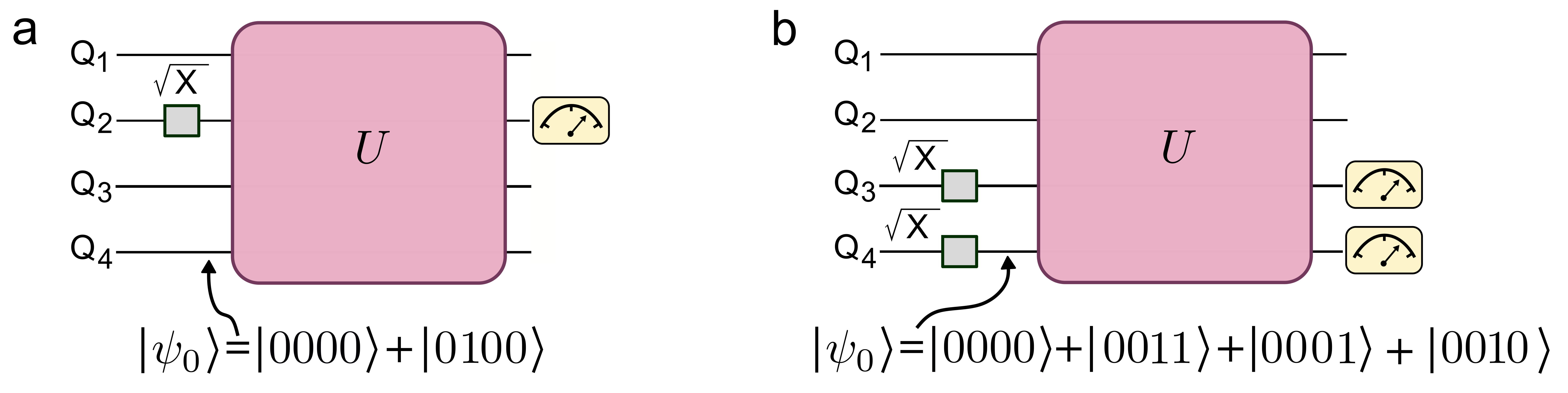}
    \caption{Pulse sequences for determining the spectra of \textbf{a,} a single excitation, and \textbf{b,} two excitations for 4 qubits. }
    \label{fig:UTexamples}
\end{figure}
The number of independent Pauli strings that need to be measured scales exponentially with the number of excitations in the manifold, and hence this method is not scalable to large photon number systems. 
\clearpage
\section{Supplementary data}

\subsection{Trajectories}
\begin{figure}[h!]
    \centering
    \includegraphics[width=0.6667\columnwidth]{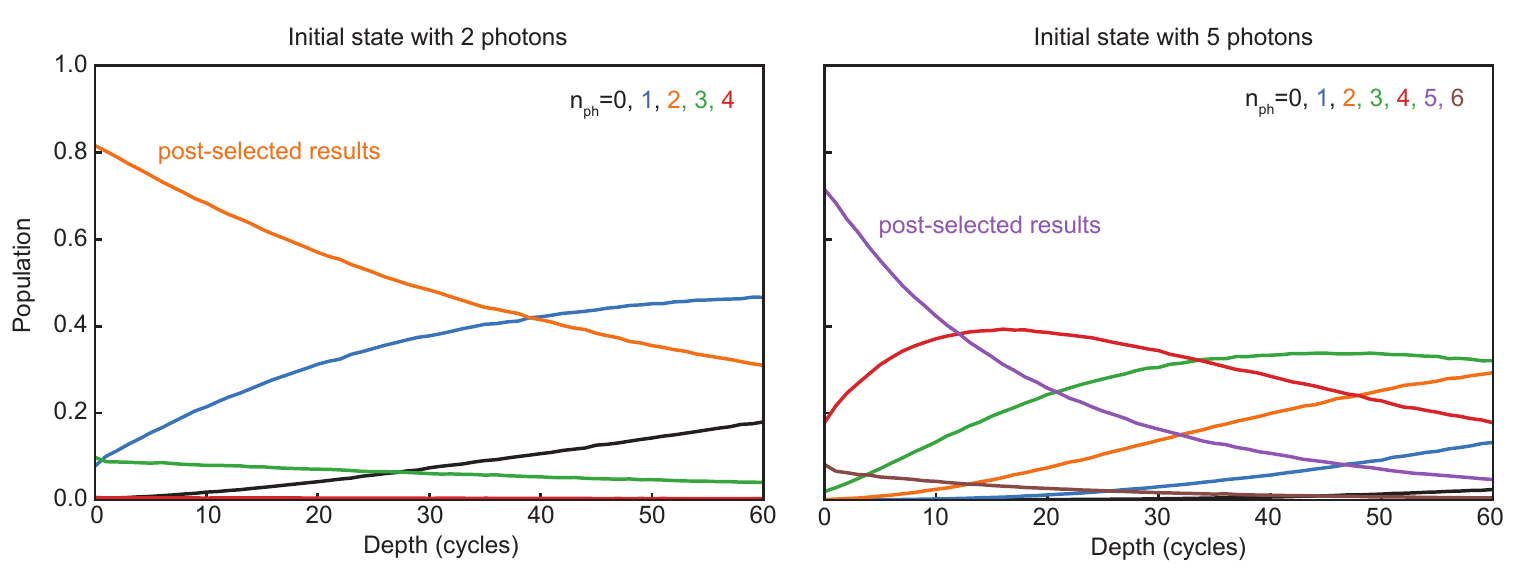}
    \caption{ \textbf{Post-selection for Figure 2:}
    In the trajectory experiments shown in the main text (Figures 2 and 4), we post select the outcome bitstrings that preserve the number of photons. This post-selection is justified by the fact that the fSim gate is an excitation preserving gate for any angle. The observed decay is due to the $T_1$ decay of the qubits during the circuit. The measurement becomes more susceptible to $T_1$ errors as the size of the bound state increases, thus increasing the necessary number of repetitions to construct the statistics.}
    \label{fig:sup_post_select}
\end{figure}

\begin{figure}[h]
    \centering
    \includegraphics{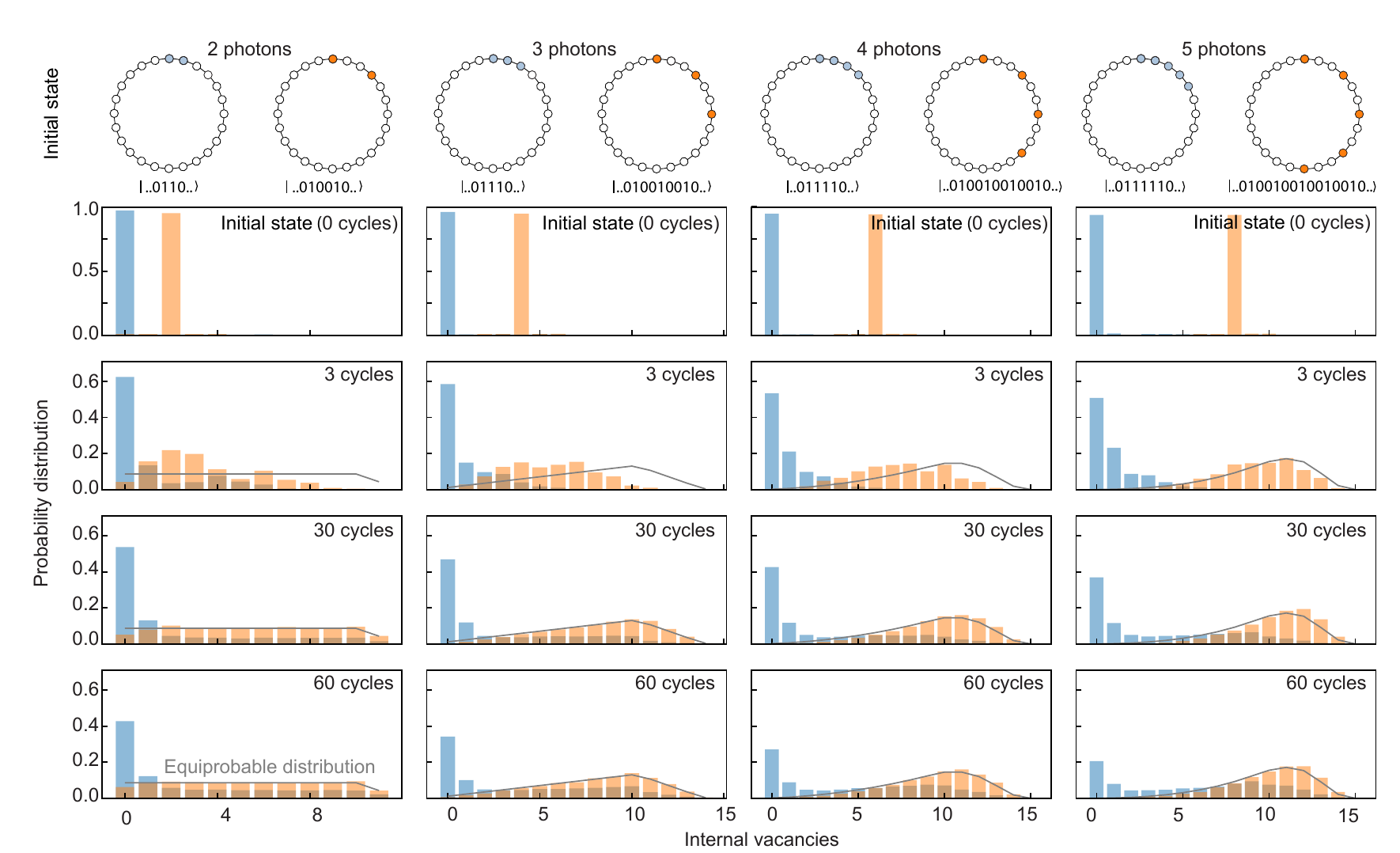}
    \caption{\textbf{Trajectory histogram:} Complementary data for Figure 2 in the main text with angles $\theta=\pi/6$ and $\phi=2\pi/3$. For each number of photons, we initialize the system with adjacent excitations (blue) or with excitations separated by a few vacancies (orange). In the case where the excitations are initially adjacent, we find that the distribution stays concentrated in the part of Hilbert space with few vacancies. When the initial state contains vacancies, on the other hand, the evolution tends toward exploring the entire Hilbert space and approaches the equiprobable distribution represented by a dashed line.
    }
    \label{fig:trajectory_hist}
\end{figure}
\clearpage

\subsection{Spectroscopy}

\begin{figure}[h]
    \centering
    \includegraphics{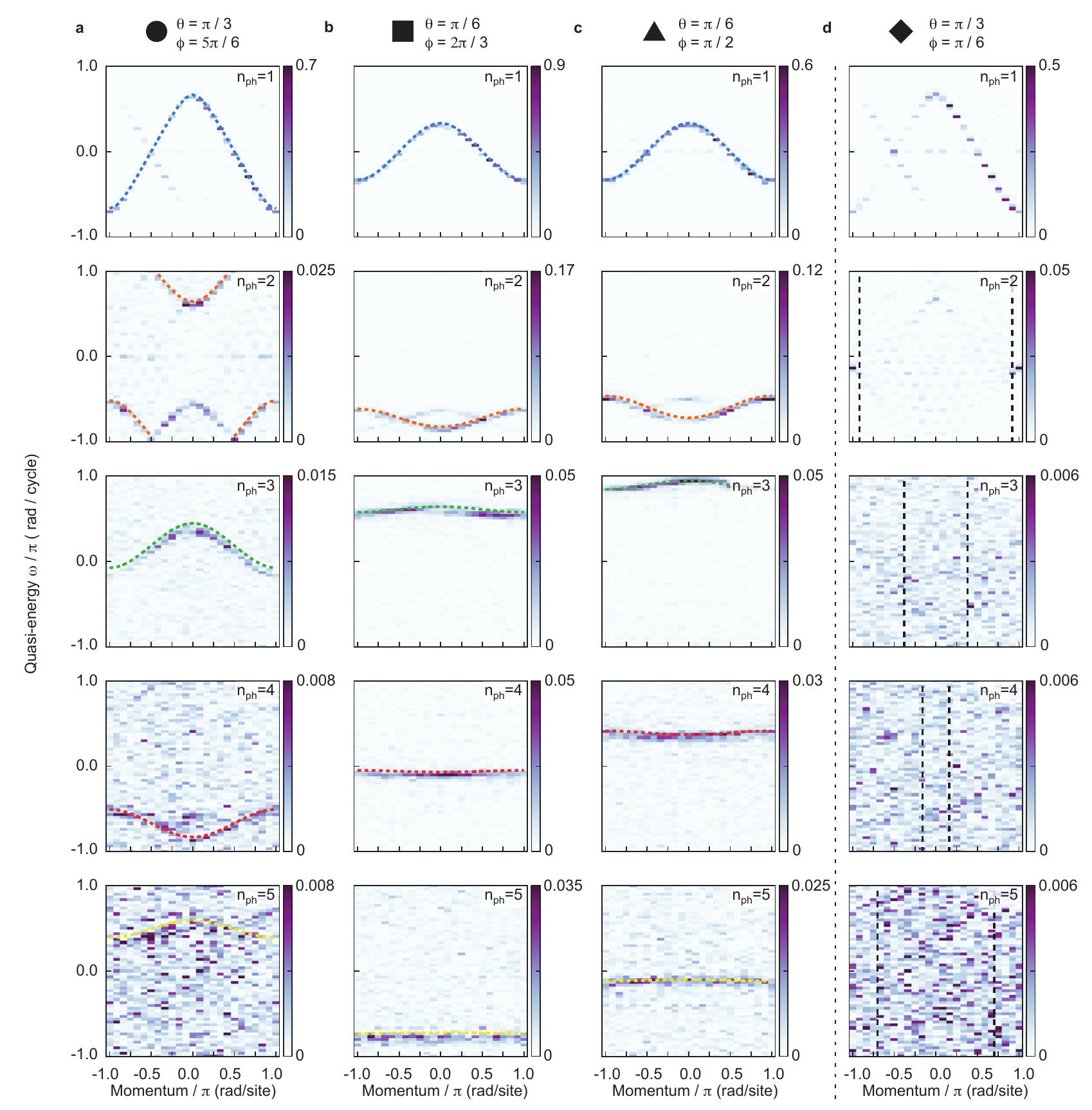}
    \caption{\textbf{Many-body spectroscopy for various fSim angles. a}, Band structures analogous to those shown in Fig. 3c in the main text for $n_{\mathrm{ph}}=1-5$ (top to bottom) and fSim angles $\theta=\pi/3$, $\phi=5\pi/6$ (\textbf{a}), $\theta=\pi/6$, $\phi=2\pi/3$ (\textbf{b}), $\theta=\pi/6$, $\phi=\pi/2$ (\textbf{c}), $\theta=\pi/3$, $\phi=\pi/6$ (\textbf{d}). Dashed lines: theoretical prediction. We observe very good agreement with the theoretically predicted band structure across all the three angle pairs that satisfy $\phi>2\theta$, required to have bound states at all momenta.  
    As predicted by theory, the bands are found to shift with increasing interaction strength ($\phi$), and the width of the dispersion increases with increasing $\theta$. 
    In the regime where $\phi<2\theta$, we observe a two-photon bound state for momenta near $k=\pm\pi$, while no bound states are observed for higher $n_{\mathrm{ph}}$ in this regime. The latter is likely due to the fact that the overlap between the initial product state and the $n_{\mathrm{ph}}$-bound state scales as $2^{-n_{\mathrm{ph}}}$, thus causing a reduction in the signal-to-noise ratio at high $n_{\mathrm{ph}}$. Dashed black vertical lines: theoretically predicted momentum threshold for the existence of bound states.}
    \label{fig:bandstructures}
\end{figure}
\newpage

\begin{figure}[h]
    \centering
    \includegraphics{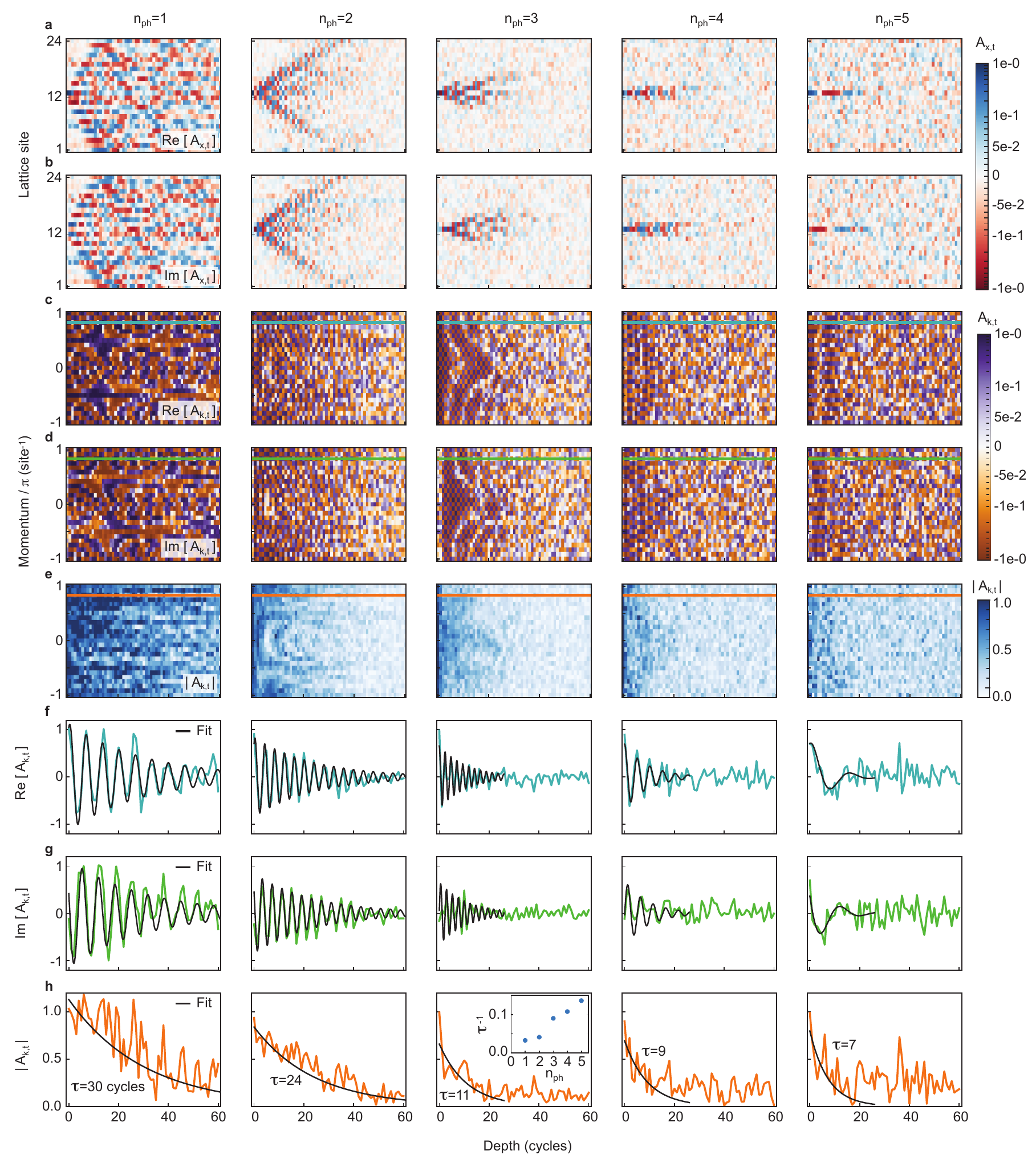}
    \caption{\textbf{Decoherence of multi-particle correlator. a,b}, Real (\textbf{a}) and imaginary (\textbf{b}) parts of the correlator $\langle C_{j,n_{\mathrm{ph}}}\rangle=\langle\Pi_{i=j}^{j+n_{\mathrm{ph}}-1}\sigma_{i}^+\rangle$ for $n_{\mathrm{ph}}=1-5$. \textbf{c,d}, Same as \textbf{a,b}, but Fourier transformed to momentum space. \textbf{e}, Absolute value of the multi-particle correlator Fourier transformed to momentum space. \textbf{f-h}, Linecuts at $k=5\pi/6$ for \textbf{c}, \textbf{d}, and \textbf{e}, respectively. Black curves show fits on the form $A_{k=5\pi/6,t}=\alpha e^{(i\omega-1/\tau)t}$ (single fit for all three plots at each photon number). Inset in center panel: extracted decay rate $\tau^{-1}$ (in units of cycles$^{-1}$) as a function of photon number.}
    \label{fig:decayfig}
\end{figure}

\newpage
\begin{figure}[h]
    \centering
    \includegraphics{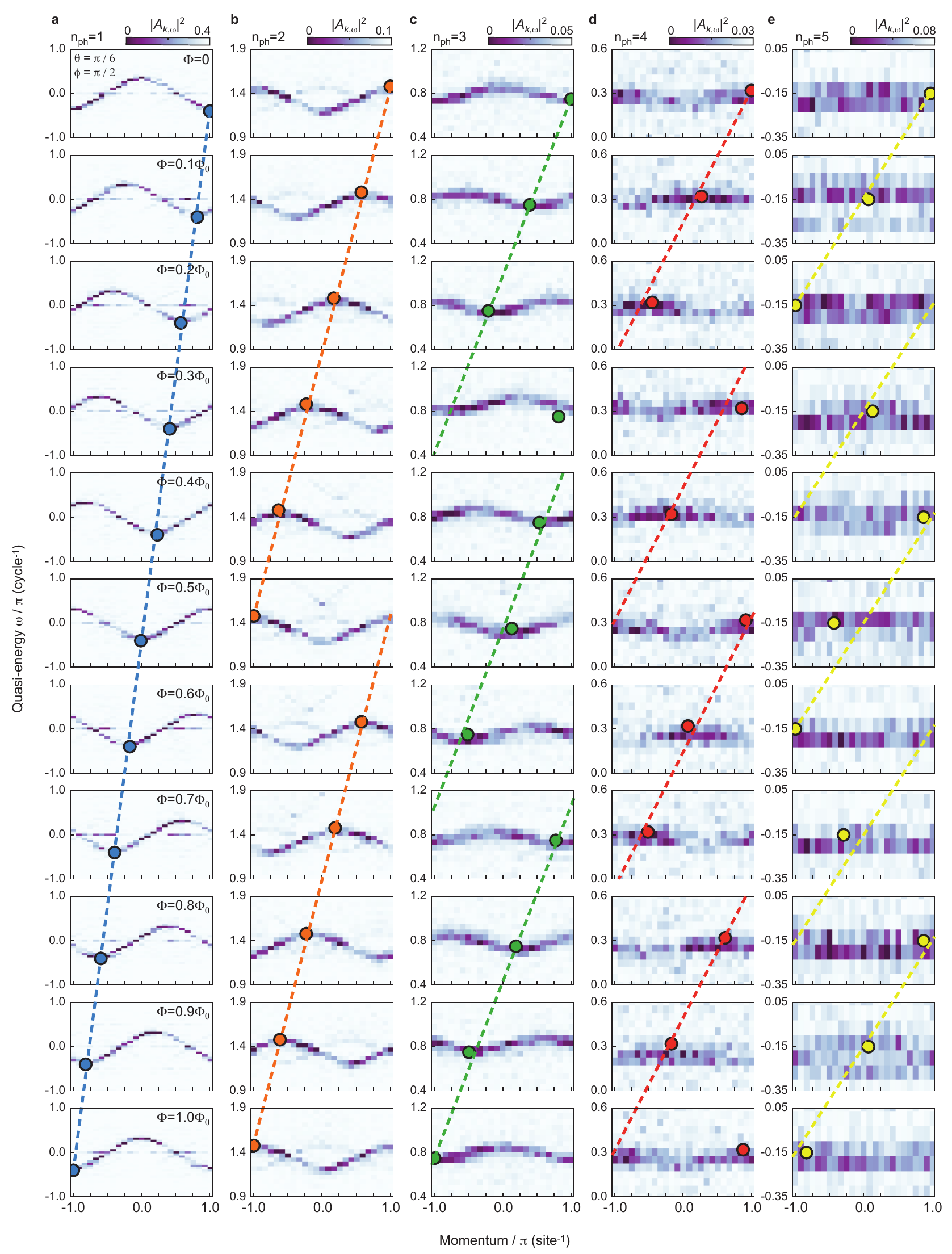}
    \caption{\textbf{Flux dependence of band structures} for $n_{\mathrm{ph}}=1-5$ (\textbf{a}-\textbf{e}, respectively). Momentum shifts were extracted by convolving the band structures with that at $\Phi=0$ ($\Phi=0.2\Phi_0$ for $n_{\mathrm{ph}}=5$ due to more clear structure), summing over the energy axis, and finding the maximum. Colored dots indicate the corresponding extracted peak positions of the bands.}
    \label{fig:fluxdep}
\end{figure}
\newpage

\begin{figure}[h]
    \centering
    \includegraphics{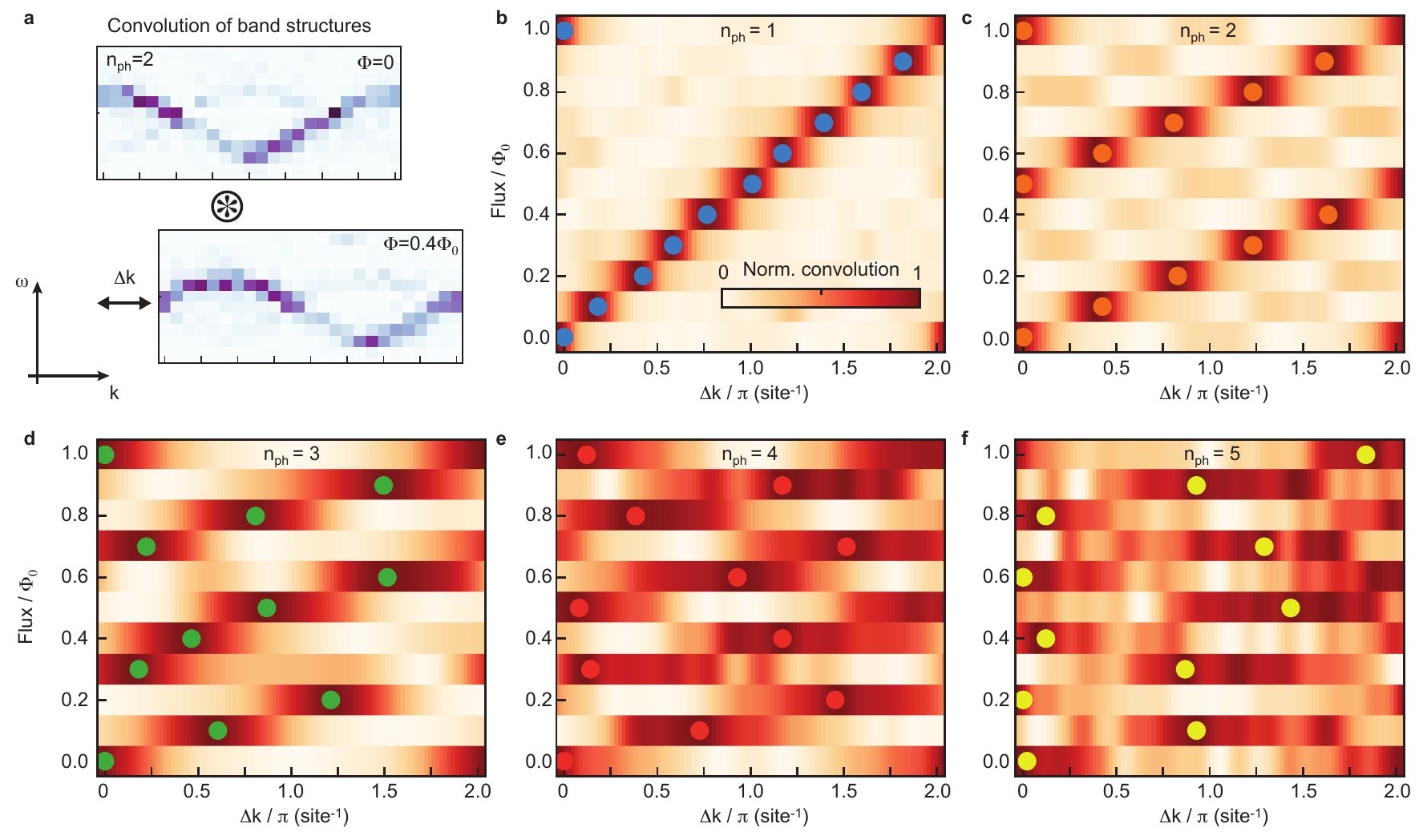}
    \caption{\textbf{Convolution technique for extracting momentum shifts. a}, The flux-induced momentum shifts shown in Fig. 3e,f in the main text are determined through cyclic convolution of the band structures ($|A_{k,\omega}|^2$) with that at $\Phi=0$ ($\Phi=0.2\Phi_0$ for $n_{\mathrm{ph}}=5$ due to more clear structure), followed by summation over the energy axis. The shift is then found from the maximum of the convolution. The signal-to-noise ratio is maximized by only including energies in the vicinity of the band. Moreover, in order to extract momentum shifts with a finer resolution than $2\pi/24$ site$^{-1}$, the band structures are interpolated along the momentum direction before convolution, increasing the number of points from 24 to 100. \textbf{b-f}, Normalized convolutions for $n_{\mathrm{ph}}=1-5$ for the full flux range ($\Phi=0-\Phi_0$). Each row is normalized relative to its respective extrema, such that their ranges span from 0 to 1. Colored dots indicate the maxima at each flux value.
    }
    \label{fig:conv}
\end{figure}
\newpage

\break
\clearpage

\vspace{1em}
\begin{flushleft}
    {\hypertarget{authorlist}{${}^\dagger$}  \small Google Quantum AI and Collaborators}

    \bigskip

    \renewcommand{\author}[2]{#1\textsuperscript{\textrm{\scriptsize #2}}}
    \renewcommand{\affiliation}[2]{\textsuperscript{\textrm{\scriptsize #1} #2} \\}
    \newcommand{\corrauthora}[2]{#1$^{\textrm{\scriptsize #2}, \hyperlink{corra}{\ddagger}}$}
    \newcommand{\corrauthorb}[2]{#1$^{\textrm{\scriptsize #2}, \hyperlink{corrb}{\mathsection}}$}
    \newcommand{\xGoogle}{\affiliation{1}{Google Research, Mountain View, CA, USA}}

\newcommand{\xGeneva}{\affiliation{2}{Department of Theoretical Physics, University of Geneva, Quai Ernest-Ansermet 30, 1205 Geneva, Switzerland}}

\newcommand{\xUMass}{\affiliation{3}{Department of Electrical and Computer Engineering, University of Massachusetts, Amherst, MA, USA}}

\newcommand{\xYale}{\affiliation{4}{Department of Applied Physics, Yale University, New Haven, CT 06520, USA}}

\newcommand{\xCaltech}{\affiliation{5}{Institute for Quantum Information and Matter, California Institute of Technology, Pasadena, CA, USA}}

\newcommand{\xUCR}{\affiliation{6}{Department of Electrical and Computer Engineering, University of California, Riverside, CA, USA}}

\newcommand{\xUCSB}{\affiliation{7}{Department of Physics, University of California, Santa Barbara, CA, USA}}

\begin{footnotesize}

\newcommand{\Google}{1}
\newcommand{\Geneva}{2}
\newcommand{\UMass}{3}
\newcommand{\Yale}{4}
\newcommand{\Caltech}{5}
\newcommand{\UCR}{6}
\newcommand{\UCSB}{7}

\corrauthora{A. Morvan}{\Google},
\corrauthora{T. I.~Andersen}{\Google},
\corrauthora{X. Mi}{\Google},
\corrauthora{C. Neill}{\Google},
\author{A. Petukhov}{\Google},
\author{K. Kechedzhi}{\Google},
\author{D. A. Abanin}{\Google,\! \Geneva},
\author{R. Acharya}{\Google},
\author{F. Arute}{\Google},
\author{K. Arya}{\Google},
\author{A. Asfaw}{\Google},
\author{J. Atalaya}{\Google},
\author{R. Babbush}{\Google},
\author{D. Bacon}{\Google},
% ...
\corrauthorb{I. Aleiner}{\Google},
\corrauthorb{L. B.~Ioffe}{\Google},
\corrauthorb{P. Roushan}{\Google}

\bigskip

\xGoogle
\xGeneva
\xUMass
\xYale
\xCaltech
\xUCR
\xUCSB

{\hypertarget{corra}{${}^\ddagger$} These authors contributed equally to this work.}\\
{\hypertarget{corrb}{${}^\mathsection$} Corresponding author: igoraleiner@google.com}\\
{\hypertarget{corrb}{${}^\mathsection$} Corresponding author: ioffel@google.com}\\
{\hypertarget{corrb}{${}^\mathsection$} Corresponding author: pedramr@google.com}
\end{footnotesize}

\end{flushleft}

\bibliography{main.bib}%